# Effect of impact velocity and acoustic fluidization on the simple-to-complex transition of lunar craters

Elizabeth A. Silber[1,2], Gordon R. Osinski[2,3], Brandon C. Johnson[1], Richard A. F. Grieve[3]

[1]Department of Earth, Environmental and Planetary Science, Brown University, Providence, RI, 02912, USA
[2]Centre for Planetary Science and Exploration / Department of Physics and Astronomy, Western University, London, Ontario, N6A 3K7, Canada
[3]Department of Earth Science, Western University, London, Ontario, N6A 3K7, Canada

Corresponding author:

Elizabeth A. Silber
Department of Earth, Environmental and Planetary Science
Brown University
324 Brook St.
Providence, RI
02912-1846
USA
E-mail: esilber [at] uwo.ca

## Abstract

We use numerical modeling to investigate the combined effects of impact velocity and acoustic fluidization on lunar craters in the simple-to-complex transition regime. To investigate the full scope of the problem, we employed the two widely adopted Block-Model of acoustic fluidization scaling assumptions (scaling block size by impactor size and scaling by coupling parameter) and compared their outcomes. Impactor size and velocity were varied, such that large/slow and small/fast impactors would produce craters of the same diameter within a suite of simulations, ranging in diameter from 10–26 km, which straddles the simple-to-complex crater transition on Moon. Our study suggests that the transition from simple to complex structures is highly sensitive to the choice of the time decay and viscosity constants in the Block-Model of acoustic fluidization. Moreover, the combination of impactor size and velocity plays a greater role than previously thought in the morphology of craters in the simple-to-complex size range. We propose that scaling of block size by impactor size is an appropriate choice for modeling simple-to-complex craters on planetary surfaces, including both varying and constant impact velocities, as the modeling results are more consistent with the observed morphology of lunar craters. This scaling suggests that the simple-to-complex transition occurs at a larger crater size, if higher impact velocities are considered, and is consistent with the observation that the simple-to-complex transition occurs at larger sizes on Mercury than Mars.

## 1. Introduction

Impact cratering is arguably the most pervasive geologic process in the solar system [e.g., *Melosh*, 1989; *Osinski and Pierazzo*, 2012]. After passage of an impact-generated shockwave and the following rarefaction wave, the residual velocity of material sets up an excavation flow. This excavation flow ultimately produces a bowl-shaped transient cavity. Although the collapse of steep crater walls leads to production of a breccia lens, small craters known as simple craters maintain a bowl shape after collapse and the final crater typically has a depth-to-diameter ratio of 1:5 [*Melosh and Ivanov*, 1999]. At larger sizes, craters undergo floor failure, leading to relatively flat floored craters with central peaks and uplifted strata near their centers [*Melosh*, 1989]. These complex craters exhibit terraced rims and their depths depend weakly on crater diameter [*Kalynn et al.*, 2013; *Clayton et al.*, 2013]. Around the simple-to-complex transition diameter are so-called transitional craters that exhibit features of both simple and complex structures (e.g., flat floors), but lack a central peak and, therefore, cannot be classified as either simple or complex. Since the transition from simple to complex structures is a function of surface gravity (g), with a roughly 1/g dependence, it occurs at different diameters on different planetary bodies [e.g., *Melosh*, 1989]. On the Moon, the simple-to-complex transition occurs at approximately 20 km [*Pike*, 1977a,b; 1980].

Figure 1 shows the progression from simple to complex craters on the Moon, which illustrates very broad morphological differences among these craters. For example, even though the average crater diameter at which the transition occurs on Moon is about 19 km, there are significant morphological differences (e.g., depth) among the craters of the same diameter [e.g., *Kalynn et al.*, 2013; *Clayton et al.*, 2013]. The explanation for this diversity of crater shapes among same-size craters is not well understood. Material properties and target parameters (e.g. damage history of the rocks, layering, porosity) play a notable role in crater morphology [*Housen and Holsapple*, 2000; *Collins et al.*, 2002; *Grieve and Therriault*, 2004; *Wünnemann et al.*, 2006; *Collins et al.*, 2011]; however, target property variations cannot account for all the observed differences in lunar transitional craters on similar terrains. Although the effect of impact velocity has been recognized as an important parameter in impact cratering [e.g., *Grieve and Cintala*, 1992; *Xiao et al.*, 2014], its influence on crater morphology and the transition from simple to complex structures has not been explored. An aim of this work is to establish the effect of impact velocity on crater morphology near the simple-to-complex crater transition.

Isolating the role of impact velocity on crater formation is not trivial, however, because of the uncertainty surrounding the physical explanation for the simple-to-complex transition and complex crater formation in general. It is well known that the formation of complex craters requires a weakening of the target rocks displaced by the impact [e.g., *Melosh*, 1977; *Melosh and Ivanov*, 1999; *Kenkmann et al*., 2013]. For instance, numerical modeling by *McKinnon* [1978] suggests that floor failure and structural uplifts only occur if the target material friction coefficient is less than 0.035, where a typical rock friction coefficient is ~0.5 – 0.7 [*Jaeger et al.*, 2009, Chapter 3]. The width of rim terraces suggests a plastic rheology, with a yield stress ~1 – 3 MPa [*Pearce and Melosh*, 1986]. Laboratory experiments of crater collapse in plasticine or clay performed by D. E. Gault produce a final crater structure strikingly similar to that of complex craters [*Melosh*, 1989]. The empirical or phenomenological evidence indicates that the Bingham plastic rheology with a yield strength of approximately 3 MPa describes the morphology of complex craters well [*Melosh*, 1977]. These strength properties are all much lower than typical values for rocks. The physical explanation for why rock would behave this way during crater collapse is not yet resolved.

Several possible weakening mechanisms have been proposed, including bulk shear strength reduction via lubrication by friction generated melt [*Dence et al.*, 1977; *Spray and Thompson*, 1995] and lubrication by impact melting [*Scott and Benn*, 2001]. Senft and Stewart [2009] and Crawford and Schultz [2013] explored temporary weakening through strain-rate dependent mechanisms along fault zones. However, the weakening mechanism most widely adopted in numerical impact simulations is acoustic fluidization [*Melosh*, 1979].

According to this idea, pressure fluctuations in the fragmented rock mass behind the impact-generated shock wave periodically allow sliding to occur at lower shear stresses than would occur under the normal overburden pressure. The space- and time-averaged result of this process provides a temporary “fluidization” of this material for as long as strong pressure fluctuations persist.

Acoustic fluidization is the most widely adopted explanation, because numerical models that employ it as a weakening mechanism have successfully reproduced many specific craters and the general crater size-morphology progression [e.g. *Wünnemann and Ivanov*, 2003; *Collins*, 2014; *Baker et al.*, 2016]. However, there are unresolved issues pertinent to assessing the effect of impact velocity on crater formation, such as how to scale the acoustic fluidization model parameters with impactor size and impact velocity. Two widely adopted acoustic fluidization scaling assumptions are to scale the intensity and duration of fluidization by impactor size [*Wünnemann and Ivanov*, 2003] and by transient crater size [*Ivanov and Artemieva*, 2002].

To explore the effect of impact velocity on crater morphology assuming acoustic fluidization is the primary transient weakening mechanism driving crater collapse, we use numerical modeling to investigate this problem, as it applies to lunar craters in the simple-to-complex regime. We compare the two commonly used acoustic fluidization scaling assumptions to quantify and contrast their effect on crater morphology and progression from simple to complex structures. The insights and results obtained in this study can be extended to transitional craters on any solid planetary body.

## 2. Scaling of Transient and Final Crater Size

Impact scaling laws [*Schmidt and Holsapple*, 1982; *Holsapple and Schmidt*, 1982; *Housen et al.*, 1983; *Holsapple and Schmidt*, 1987; *Schmidt and Housen*, 1987; *Holsapple*, 1993] based on laboratory scale impact experiments were developed with an aim to quantify the relationship between various impact parameters and the size of the transient cavity. Separate scaling laws, based on detailed observations of craters, provide an estimate of the final crater resulting from a transient cavity of a given size (see, for example, *Holsapple and Schmidt* [1987], *Holsapple* [1993] *Ivanov and Artemieva* [2002], and *Johnson et al.* [2016a], who compare several independently derived scaling laws). In addition to providing an estimate of the outcome of a given impact, these scaling laws are useful for testing the numerical models.

An important consideration is the late-stage equivalence principle [*Huang and Chou*, 1968; *Billingsley*, 1969; *Dienes and Walsh*, 1970]. Developed from blast wave theory, a similarity concept or "late-stage equivalence", indicates that at some point in time (or space), the details of the projectile will no longer influence the terminal effects of the impact; in this regard, the impact is equivalent to a point source of energy and momentum [*Taylor*, 1950; *Sedov*, 1959; *Sakurai*, 1964].

Based upon the principles of the late-stage equivalence, Holsapple and Schmidt [1987] characterized the coupling parameter ($C$) [*Holsapple*, 1981, 1983], to describe the coupling of the impactor energy and momentum into the target:

$$C = D_i v_i^{\mu} \rho^{\nu} \qquad (1).$$

Here, $D_i$ is the impactor diameter, $v_i$ is the impactor velocity and $\rho$ is density. This approximation, however, falls somewhere between the kinetic energy and momentum regimes ($\nu = 1/3$, $1/3 \leq \mu \leq 2/3$). For non-porous materials (e.g., competent rock), the value of $\mu$ is ~0.55, as found in numerous experiments [*Housen and Holsapple*, 2011]. Thus, it follows that all impacts with equal $C$ (where $D_i$ and $v_i$ take some realistic value) are expected to produce a transient cavity of the same size (note that transient cavity diameter does not scale linearly with coupling parameter). However, the problem is much more complex than can be presented here and the interested reader is directed to Holsapple and Schmidt [1987] for the full discussion.

## 3. The Block-Model of Acoustic Fluidization and Scaling of Model Parameters

Although recent work has made significant progress toward implementation of the original description of acoustic fluidization in a shock physics code [*Hay et al*., 2014], a simplified model (the Block-Model) of acoustic fluidization [*Ivanov and Kostuchenko*, 1997; *Melosh and Ivanov*, 1999] has tended to be adopted in impact simulations.

In the Block-Model of acoustic fluidization, some fraction of strong, transient pressure fluctuations (seismic energy) initiated by the passage of the impact-generated shock wave is responsible for temporarily counteracting overburden pressure, thereby reducing the frictional resistance of the blocks within granular breccia. In iSALE shock physics code, the vibrational pressure ($P_{vib}$) is calculated from the maximum vibrational particle velocity ($v_{vib}$) through:

$$P_{vib} = \rho c_s v_{vib} \quad (2),$$

where $\rho$ and $c_s$ are the bulk density and sound speed of the cell, respectively [*Ivanov and Turtle*, 2001; *Wünnemann*, 2001 (Chap. 3.5, Eq. 3.19, p. 91)]. The vibrational velocity is assumed to be some fraction (typically 10%) of the magnitude of the particle velocity behind the shockwave, up to some maximum velocity as defined by the user (here 200 m/s). After passage of the shockwave, the vibrational velocity is decreased according to an exponential decay law [*Ivanov and Kostuchenko*, 1997; *Melosh and Ivanov*, 1999; *Collins et al.*, 2002], with a characteristic decay time constant $T_{dec}$. The vibrational pressure acts to reduce the effective pressure employed in the strength model; in addition, the strength is augmented by a rate-dependent term, scaled by an effective viscosity of the acoustically fluidized material $\eta_{lim}$ [*Melosh and Ivanov*, 1999; *Ivanov and Turtle*, 2001]. For example, in the simplified situation where the static strength $Y$ is simply directly proportional to pressure $P$, $Y=\mu P$ (where $\mu$ is the coefficient of friction), the effective strength in the presence of vibrations becomes $Y_{vib} = \mu(P - P_{vib}) + \eta_{lim}\rho\dot{\epsilon}$, where $\rho$ is density and $\dot{\epsilon}$ is the invariant deviatoric strain rate [*Melosh and Ivanov*, 1999; *Ivanov and Turtle*, 2001]. If the viscous vibrational strength is greater than the static strength, the latter is used so that acoustic fluidization acts only to reduce friction. For a more detailed overview, the reader is directed to Melosh and Ivanov [1999], Ivanov and Turtle [2001], Collins et al. [2002] and Wünnemann and Ivanov [2003].

To replicate a specific impact event, the two free Block-Model parameters that control the weakening process, the kinematic viscosity of the fluidized region ($\nu_{lim}$) and the decay time of the block vibrations ($T_{dec}$), must be specified [e.g., Collins et al., 2002]. To replicate impacts at all sizes, and in particular phenomena at the simple to complex transition, rationale have been developed to describe how $T_{dec}$ and $v_{lim}$ scale with impact event size [*Ivanov and Artemieva*, 2002; *Wünnemann and Ivanov*, 2003; *Bray et al*., 2014].

The conceptual premise of this scaling is that the target is represented by a system of large, discrete blocks (comprised of shattered target rock), each of characteristic size $h$, that oscillate at some period ($T$) within a matrix of smaller fragments. In this case, $T_{dec}$ and $v_{lim}$ can be related to the block size and period [*Ivanov and Artemieva*, 2002]. In a situation of strong vibrations, the motion of the completely fluidized material can be described as a viscous motion with an effective kinematic viscosity:

$$v_{lim} = c_{af}\, h^2/T \qquad (3),$$

were $c_{af}$ is a numerical coefficient with values from 4 to 8, depending on the model assumptions [*Ivanov and Artemieva*, 2002]. The block oscillation decay time ($T_{dec}$) is closely related to the quality factor ($Q$), which is the ratio of the energy stored to the energy lost (per cycle):

$$T_{dec} = QT \qquad (4).$$

Thus, Ivanov and Artemieva [2002] proposed that the period of oscillations is controlled by the matrix (or soft breccia, with density $\rho_b$, thickness $h_b$, characteristic sound speed $c_b$, and compressibility $\rho_b c_b^2$), which dampens the block (height $h$ and density $\rho$) movement. Using a relation for simple harmonic oscillations they derive expressions for $T_{dec}$ and $v_{lim}$:

$$v_{lim} = c_b\, h/[(\rho/\rho_b)(h_b/h)]^{½} \qquad (5)$$

and

$$T_{dec} = 2\pi\, Q\, h/c_b\, [(\rho/\rho_b)(h_b/h)]^{½} \qquad (6).$$

According to this rationale, if $Q$, $c_b$ and $(\rho/\rho_b)(h_b/h)$ are constant, then the characteristic oscillation period ($T$) is proportional to block size ($h$), implying that both $v_{lim}$ and $T_{dec}$ scale linearly with block size. In other words, all that remains to specify the scaling of the Block-Model parameters is to determine how the characteristic block size $h$ scales with impact size.

Numerous numerical studies have shown that to match the progressive change in crater morphology with size, the block size must be some function of impact event size [e.g. *Ivanov and Artemieva*, 2002; *Wünnemann and Ivanov*, 2003; *Bray et al*., 2014]. However, direct measurements of characteristic block size are rare. The core drilling at the Puchezh-Katunki impact structure in Russia revealed that the block size beneath the 40 km diameter crater is ~100 m [*Ivanov et al.*, 1996]. Block sizes ranging from 50 – 100 m were ascertained through the geological mapping of impact structures, 7 km and 6 km in diameter, respectively, at Upheaval Dome, USA [*Kenkmann et al*., 2006] and Waqf as Suwwan, Jordan [*Kenkmann et al*., 2010]. The observations were consistent with increase in block size as a function of distance from the crater center [*Kenkmann et al*., 2012]. On the other hand, observations at West Clearwater Lake show a much more variable block size (< 1 m – ~43 m), and as such do not fit the block/breccia template [*Rae et al*., 2017]. Before general assumptions can be made, however, it would be necessary to conduct more field observations. In the meantime, numerical modeling when compared to the observed morphometry of craters on planetary surfaces remains the primary mode of inferring complex and difficult to directly observe elements of the cratering process.

Ivanov and Artemieva [2002] proposed that block size scales linearly with transient crater size and hence that $v_{lim}$ and $T_{dec}$ are invariant for all impact scenarios that produce the same size transient crater. On the other hand, Wünnemann and Ivanov [2003] proposed that the block size might scale linearly with impactor size such that:

$$v_{lim} = \gamma_\eta c_b R_i \quad (7)$$

and

$$T_{dec} = \gamma_\beta (R_i/c_b) \quad (8).$$

Here, $R_i$ is the impactor radius, and $\gamma_\eta$ and $\gamma_\beta$ are the viscosity and time decay acoustic fluidization constants, which serve as model inputs in iSALE. Considerable success with this approach has been achieved by deriving the acoustic fluidization constants ($\gamma_\eta$ and $\gamma_\beta$) empirically by matching modelling results to actual crater dimensions and/or morphology [e.g., *Wünnemann and Ivanov*, 2003; *Collins*, 2014; *Milbury et al.*, 2015; *Baker et al*., 2016]. We note that in almost all cases the block size is assumed not to vary in space and/or time during the simulation.

In principle, a linear scaling between block size and impactor radius, implies that large impactors will produce larger block fragments, and consequently a longer vibration decay time and a higher effective viscosity than would smaller impactors. Wünnemann and Ivanov [2003] indicated that while this scaling approach would be appropriate for a regime where the impact velocity can be assumed to be relatively constant and other target parameters negligible, it is not meant to be applicable across different velocity regimes.

For example, Wünnemann and Ivanov [2003] varied the values of acoustic fluidization constants ($\gamma_\eta$ = 0.1 – 0.8 and $\gamma_\beta$ = 150 – 400) at constant impact velocity ($v_i$ = 15 km/s) for a range of impactor sizes to replicate the depth-diameter dependence with crater size in an acoustically fluidized target. While they successfully replicated the simple-to-complex transition behaviour, they note that the acoustic fluidization parameters set appropriate for the Moon might not be applicable to planetary bodies where the average impactor velocity might be significantly different (e.g. Mercury). However, in numerical modeling studies, it is common practice to use invariant values for $\gamma_\eta$ and $\gamma_\beta$ over a range of impact sizes, whether it is for impacts occurring at some constant velocity [e.g., *Wünnemann and Ivanov*, 2003; *Collins*, 2014; *Baker et al.*, 2016] or a range of impact velocities [e.g., *Miljković et al.*, 2013].

To demonstrate the differences between scaling only by impactor size [*Wünnemann and Ivanov*, 2003] or by the transient cavity diameter [*Ivanov and Artemieva*, 2002], it is helpful to consider a combination of impactor size and velocity that produce the same size transient cavity (e.g., small/fast vs. large/slow impactor). To briefly recap, before we discuss these two approaches in more detail, according to the impactor size scaling, a small (and fast) impactor that produces the same transient cavity diameter as the large (and slow) impactor will result in smaller block size, shorter decay time and lower viscosity. This is in direct contradiction with the transient cavity diameter scaling of block size, which advocates that the block size will always be the same, regardless of the impactor size and velocity combination, as long as the resulting transient cavity is of the same diameter.

The coupling parameter can be applied to compute the impactor sizes corresponding to impact velocities of interest, as such combinations would lead to a transient cavity of the same size (and consequently the same block size). Since $T_{dec}$ and $\eta_{lim}$ are assumed to remain invariant for a given size crater, then the final step is to utilize equations (7) and (8) and derive $\gamma_\eta$ and $\gamma_\beta$ for any impactor (and, thus, impact velocity) for a particular crater. We refer to this approach as coupling parameter scaling from here on.

The scaling by coupling parameter should satisfy the late-stage equivalence principle. In far field (e.g., far from the point of impact), the shock wave, along with the rarefaction wave should remain the same for a given transient crater size. It then follows that block size, $T_{dec}$ and $v_{lim}$ are also invariant for some specified crater diameter. Thus, we would expect slow/large and fast/small impactors to produce craters with similar morphologies. Note, however, that the particle velocity associated with fast vs. slow impact velocity will be different in the near field.

On the other hand, the scaling by the impactor size employs a very different approximation. The acoustic field will exhibit a disparity between craters formed by large/slow and small/fast impactors. Thus, in the near-field (e.g. close to the point of origin), the shock wave will not be the same for various impactor size/velocity combinations; large and slow impactors will produce larger blocks (fragments), as opposed to small and fast impactors which will generate comparatively small blocks. This assumption is not compatible with the late-stage equivalence principle for scenarios where acoustic fluidization is important. The effect of acoustic fluidization will last much longer and have more influence on crater collapse in craters produced by large and slow impactors. Conversely, small and fast impactors will produce notably shorter lasting acoustic vibrations field, thereby halting the crater collapse. These two dramatically different outcomes are expected to significantly affect the crater morphologies for a given transient crater size. While scaling by impactor size might appear more intuitive, it is imperative to compare these two scaling approaches side by side.

Additionally, in modeling studies, the acoustic fluidization parameters (whether expressed as $T_{dec}$ and $v_{lim}$, or $\gamma_\eta$ and $\gamma_\beta$), are often assumed invariant across varying impact velocities [e.g., *Miljković et al.*, 2013; *Bray and Schenk*, 2015], while others keep the impact velocity constant to avoid the issue of scaling [e.g., *Wünnemann and Ivanov*, 2003; *Collins*, 2014; *Baker et al.*, 2016]. Thus, this also motivates a comparative study of the two scaling approaches.

Although both acoustic fluidization scaling approaches are relatively crude parameterizations of the actual fragmentation process which in nature controls block sizes, at the moment these remain the most widely adopted models in hydrocode modeling of impact craters. On the other hand, Bray et al. [2014] showed that a better fit to the size-morphometry progression of craters on Ganymede could be achieved using a non-linear, as opposed to linear, scaling between block size (and breccia sound speed) and impactor size.

In Section 6, we will examine and discuss the outcomes and implications of these two acoustic fluidization scaling approaches. We now turn to model setup and numerical simulations.

## 4. Model Setup and Numerical Simulations

Simulations were carried out using the two-dimensional iSALE shock physics code [*Wünnemann et al.*, 2006], a multi-material, multi-rheology [*Melosh et al.*, 1992; *Ivanov et al.*, 1997] extension of the finite difference SALE hydrocode [*Amsden et al.*, 1980]. iSALE utilizes the material strength [*Collins et al.*, 2004], damage [*Ivanov et al.*, 2010] and porosity compaction [*Wünnemann et al.*, 2006; *Collins et al.*, 2011] models, although the latter are not used in this work. iSALE has been benchmarked against laboratory experiments and other hydrocodes [*Pierazzo et al.*, 2008], and has been used extensively to model impact cratering processes, at all scales [e.g., *Ivanov and Artemieva*, 2002; *Wünnemann and Ivanov*, 2003; *Collins and Wünnemann*, 2005; *Collins et al.*, 2008; *Potter et al.*, 2012; *Yue et al.*, 2013; *Melosh et al.*, 2013; *Collins*, 2014; *Baker et al.*, 2016].

The target and impactor were represented with ANEOS derived equations of state (EOS) for granite [*Pierazzo et al.*, 1997] and dunite [*Benz et al.*, 1989], respectively. The model strength parameters are given in Table 1. Granite is often used as a close analogue to the lunar crust [e.g., *Yue et al.*, 2013], while dunite is a reasonable approximation for typical ordinary chondrite asteroidal material [*Pierazzo et al.*, 1998; *Yue et al.*, 2013; *Svetsov and Shuvalov*, 2015]. The lunar gravity was set to 1.62 $m/s^2$. iSALE includes the material strength [*Collins et al.*, 2004] and damage [*Ivanov et al.*, 1997] models for geological materials, as well as the Block-Model [*Ivanov and Kostuchenko*, 1997; *Melosh and Ivanov*, 1999; *Ivanov and Turtle*, 2001; *Wünnemann and Ivanov*, 2003] of acoustic fluidization.

The impact velocity and the impactor size were varied in all the simulations, as these two parameters are critical with respect to the acoustic fluidization scaling choice. All other parameters, not including acoustic fluidization constants, were kept constant (e.g., target properties). This is the key aspect, as by keeping all target parameters constant, it is possible to investigate the combined effect of the impact velocity and acoustic fluidization. This approach would be appropriate for any scenario where the target properties are kept constant, regardless of the type of target involved (e.g. granular material, layered media).

We model vertical impacts with velocities of 6, 10, 15, and 20 km/s to account for a range of lunar encounter velocities [*Le Feuvre and Wieczorek*, 2011; *Yue et al.*, 2013]. We limit the highest impact velocity to 20 km/s because high impact speeds require significant computational resources and can add weeks and even months to each simulation and only ~20% of lunar impacts will occur at higher velocities [*Le Feuvre and Wieczorek*, 2011; *Yue et al.*, 2013].

The simulations were divided into four sets, where each set represents impacts resulting in the same transient cavity diameter ($D_{tr}$). That means that in one set, the varying combinations of impactor size and velocity, from large/slow on one end of the spectrum, to small/fast on the other end, will produce a crater with the same diameter (applies to both transient and final crater size). To derive the impactor sizes appropriate for given velocities, we applied the scaling law using the coupling parameter *C* (equation 1), with $\mu$ = 0.55 (in all simulations except two sets, where $\mu$ = 0.56, see Table 2) [*Housen and Holsapple*, 2011], as a starting point. This approach produced transient cavities of approximately the same diameter within a simulation set. It should be noted, while the value of $\mu$ used in our simulations is that for competent rock, as determined through laboratory experiments, its value depends on material parameters, such as friction. However, to avoid inclusion of too many unknowns into the model, we implement the above value of $\mu$ as it is representative of the problem at hand. The impactor sizes (Table 2) were chosen such that the transient cavities are 7 – 17 km in diameter corresponding to final craters that are 10 – 26 km in diameter.

Generally, the expanding shock wave damages the intact material long before the crater opens up. However, if the model is set up such that the damaged zone is smaller than the acoustically fluidized region, then the resulting crater morphologies might not be correctly predicted; hence, caution should be exercised in the model setup stage. Note that the scale of the damaged zone as compared to the zone of acoustically fluidized material likely depends on the strength and damage model employed. In their study, Wünnemann and Ivanov [2003] assumed that the entire target was fully damaged (prior to the impact). In our work, the acoustically fluidized region is significantly smaller than the damaged zone for all simulations.

Recent studies have demonstrated that the acoustic fluidization constants, $\gamma_\beta$ = 300 and $\gamma_\eta$ = 0.015, are appropriate good choice at impact velocity of 15 km/s, resulting in good agreement between simulated and observed morphology over a range of crater sizes [*Collins*, 2014; *Baker et al.*, 2016]. Hence, we use these values as the starting point in our simulations.

Each set of simulations consisted of two subsets, featuring the acoustic fluidization scaling according to either the coupling parameter [*Ivanov and Artemieva*, 2002] or the impactor size [*Wünnemann and Ivanov*, 2003]. In the impactor size scaling, all simulations used constant value for $\gamma_\eta$ and $\gamma_\beta$ ($\gamma_\beta$ = 300 and $\gamma_\eta$ = 0.015), regardless of the impactor diameter or velocity. Therefore, a large and slow impactor will result in longer oscillation decay time ($T_{dec}$) and higher viscosity ($\eta$), as opposed to its small and fast conjugate (Table 2). The coupling parameter scaling, on the other hand, implies invariant $T_{dec}$ and $v_{lim}$. To set up the simulations, the following approach was applied. Since the acoustic fluidization constants ($\gamma_\eta$ and $\gamma_\beta$) and, therefore, the viscosity ($\eta$) and oscillation decay time ($T_{dec}$), are known for an impact at velocity of 15 km/s, the simple linear relations (equations 7 and 8) were then applied to derive the acoustic fluidization constants for impactor sizes at any other impact velocity within the simulation set. The resulting constants are $\gamma_\beta$ = 180, $\gamma_\eta$ = 0.00897 (6 km/s); $\gamma_\beta$ = 239, $\gamma_\eta$ = 0.0119 (10 km/s); $\gamma_\beta$ = 300, $\gamma_\eta$ = 0.015 (15 km/s), and $\gamma_\beta$ = 352, $\gamma_\eta$ = 0.0176 (20 km/s). The $T_{dec}$ and $v_{lim}$ for all simulations are listed in Table 2.

The resolution was set to 10 cells per projectile radius (CPPR) for all simulations, as it offers the best compromise between computing time and accuracy [*Wünnemann et al.*, 2006; *Wünnemann et al.*, 2008; *Pierazzo et al.*, 2008; *Elbeshausen et al.*, 2009]. We performed subsequent testing against simulations at 20 CPPR, which revealed no notable changes in crater morphology. To retain 10 CPPR resolution across all simulations, the cell size was varied depending on the impactor size. Since small/fast impactors produce proportionally larger craters, these are, thus, resolved with about twice as many cells compared to their large/slow counterparts. The mesh size for each simulation was optimized, such that there are at least 2–3 crater radii vertically and laterally in the high resolution zone.

We also compared the simulated craters to the observed lunar craters. Representative lunar crater profiles, based on a number of profiles at different azimuths, were taken from the Lunar Reconnaissance Orbiter (LRO) topographic map database (http://target.lroc.asu.edu/q3/) [*Losiak et al.*, 2009]. The profiles were taken from craters that have been defined as fresh (dating to the Eratostenian or Copernican era) [*Losiak et al.*, 2009; *Kalynn et al.*, 2013].

## 5. Results

Table 2 provides a summary of the results of our simulations. The outputs for the suite of simulations using the coupling parameter scaling are listed in the upper half of Table 2, while the outputs using the impactor size scaling are shown in the lower half of the table. Each simulation subset consisted of a combination of the impactor diameter and velocity, such that they all produced the transient cavity with less than 6% difference diameter within a simulation set. This difference of 6% is negligible and has no effect on the final results. The likely source, however, is the combined effect of the model deviation from the scaling laws and the simulation mesh resolution. The final crater diameters are also nearly identical, within < 4.5% difference (Table 2). The cell dimensions vary from 21 m to 122 m, depending on the impactor size. The transient cavity depth was measured relative to the pre-impact surface level. The final crater depth was measured from the highest point on the rim to the deepest point on the crater floor ($d_{f(max)}$ in Table 2).

The maximum depth of the transient cavity is reached within several seconds from the moment the impactor comes in contact with the target. However, the transient crater continues to expand laterally up until the end of the excavation stage, which is the point at which the crater volume reaches its first maximum [*Elbeshausen et al.*, 2009], and the ejecta curtain starts to 'kink'. During this time, in larger craters (transitional to complex), the floor will also undergo uplifting; thus, in most cases, the transient crater depth is not the same as the maximum attained depth. To avoid any ambiguity, all the transient cavity depth ($d$) and diameter ($D$) measurements were taken at the point at which the cavity volume reached its first maximum.

The transient cavity depth to diameter ($d/D$) ratio versus impact velocity does not reveal a notable trend. In both acoustic fluidization scaling scenarios, the transient cavity $d/D$ trend is nearly zero (the slope is $-2.31\cdot10^{-5}$ for the coupling parameter scaling, and $1.37\cdot10^{-3}$ for the impactor size scaling). A hint of increasing *d/D* trend versus impact velocity in impactor size scaling is likely the result of the decreasing effective viscosity of acoustic fluidization.

The transient crater depth versus diameter for all simulated craters is shown in Figure 2a. Both scaling approaches follow approximately the same trend. However, there is an indication of a slight divergence at larger crater sizes ($D_t$ > 15 km), where the coupling parameter scaling appears to result in slightly shallower craters. Without running more simulations for larger crater sizes, it is not possible to determine with much certainty if this deviation is due to the effect of acoustic fluidization (we will touch on this again in the Discussion section).

The final crater depths and diameters as measured at the end of each simulation were plotted against the observed lunar transitional and complex crater data [*Kalynn et al.*, 2013], as shown in Figure 2b. The coupling parameter scaling produces systematically deeper craters, along the upper bounds of the observed depths. Nevertheless, our simulation results are consistent with the measurements obtained by Kalynn et al. [2013] and lie within their error bars.

At this point, we further examine if the disparity in acoustic fluidization scaling models is large enough to make a notable effect on the crater morphology. The results of our simulations are presented in the time series panel plot figures. All figures are organized such that the left (panels a-c) and right (panels d-e) columns show the contrast between the end members (6 km/s and 20 km/s) for the coupling parameter scaling and the impactor size scaling, respectively. The two panels across the top row (panels a, d) show the transient cavity, the middle row (panels b, e) depicts the point in time at which the relative difference among the end members in crater depths and morphologies is the greatest, and the bottom row (panels c, f) shows the crater cross-section closer to the end of the simulation.

The morphological differences among simple craters are minor (Figure 3), regardless of the acoustic fluidization scaling choice. The point of maximum difference in crater depth for large/slow and small/fast impactors is at 190 s for coupling parameter scaling and 90 s for the impactor size scaling (Figure 3b, d). For all other craters in our simulations, the two acoustic fluidization scaling methodologies lead to appreciably different results in terms of crater temporal evolution and morphology. We show the crater cross-section as a function of time after the impact for simulation sets 200/2000 ($D_f$ = 13 km) (Figure 4), 300/3000 ($D_f$ = 20 km) (Figure

5) and 400/4000 ($D_f$ = 26 km) (Figure 6), to contrast the outcomes of the two acoustic fluidization scaling regimes, as well as the different impactor size/velocity combinations. These simulation sets exhibit large diversity in crater depths, as well as morphologies (Table 2, Figures 4 – 6). For example, in simulation sets 200/2000 ($D_f$ = 13 km), the difference in crater depths during the early phase of the crater modification stage (t = 160 s), is as much as ~1 km (impactor size scaling) and ~0.5 km (coupling parameter scaling). While the final result of simulations 200 ($v_i$ = 6 km/s) and 201 ($v_i$ = 10 km/s) is a simple crater (see Table 2 for crater classifications), higher velocity impacts (202 and 203, $v_i$ = 15 and 20 km/s) produce a crater exhibiting transitional features. This trend is contrary to that seen in simulation set 2000 ($D_f$ = 13 km), where all simulations, except the highest velocity one (2030, $v_i$ = 20 km/s), result in transitional craters (Table 2).

The depth difference is even greater in simulation sets 300/3000 ($D_f$ = 20 km), where the difference (at t = 195 s) is ~1.9 km (impactor size scaling) and ~1 km (coupling parameter scaling). Interestingly, the outcome of the coupling parameter scaling for large/slow impactors closely resembles that of the impactor size scaling for small/fast impactors, almost like a "mirror image". For example, in terms of depth and crater floor morphology, the aftermath of a 20 km/s impact with the impactor size scaling closely matches that of a 6 km/s impact with the coupling parameter scaling. This trend is particularly pronounced in simulation sets 300/3000 and 400/4000, though it is also evident in simulation sets 200/2000 ($D_f$ = 13 km) (e.g., Figure 4).

The comparison between the simulated craters and the observed lunar craters is shown in Figure 7. We plotted the crater profiles for the lowest (6 km/s) and highest (20 km/s) impact velocities for all simulation sets and compared these to the observed lunar craters. All lunar craters are listed in the figure legend – the details pertaining to these craters, such as the coordinates, are available from LPI (http://www.lpi.usra.edu/lunar/surface/). Some of these craters are also shown in Figure 1.

We also plotted the strength of the acoustic fluidization field ($Y_{Ac}$), represented by the ratio of the material strength with the acoustic fluidization included ($Y_d$) to the material strength without the acoustic fluidization ($Y_s$), as: $Y_{Ac}$ = 1-($Y_d$/$Y_s$) (Figure 8). The plots show the simulation sets 300/303 and 3000/3030 side by side, at t = 5, 80 and 120 s ($D_f$ = 20 km, $v_i$ = 6 and 20 km/s). This direct, step-by-step comparison between the simulations with two acoustic fluidization scaling modes shows that the acoustic fluidization induced by large/slow impactors will continue to drive the change in crater morphology long after the effect of acoustic fluidization has ceased for small/fast impactors in the impactor size scaling.

## 6. Discussion

The results of our study suggest that the coupling parameter and impactor size scaling methodologies lead to notable differences in the temporal evolution and morphology of the resulting craters in the simple-to-complex regime. However, in terms of overall crater morphology, simple craters ($D_f$ = 10 km) are insensitive to the acoustic fluidization scaling choice. This is not surprising, because in small craters, the effect of acoustic fluidization will not be sustained as long as in larger craters, due to relatively short $T_{dec}$.

There are two concurrent effects at work: the time decay of acoustic vibrations and the kinematic viscosity. The impactor size scaling, as opposed to coupling parameter scaling, leads to the higher values for the oscillation time decay and kinematic viscosity for the impacts produced by large/slow projectiles as opposed to small/fast projectiles. Therefore, the acoustic fluidization vibrations will continue to aid the collapse and drive the change in morphology of craters produced by large/slow impactors for a longer time than those formed by small/fast impactors, despite the fact that both scenarios will produce the craters of approximately same diameter (Figure 8). Another aspect of the two scaling considerations is the block size – according to the impactor size scaling, even if the transient cavity diameter is the same, large/slow impactors will produce larger blocks than small/fast impactors.

The viscosity will affect the transient cavity depth, also noted by Wünnemann and Ivanov [2003]. This is expected, as higher viscosities associated with large/slow impactors will suppress the flow of the material and, therefore, transient cavity growth and collapse. The absence of any strong trend in transient cavity $d/D$ can be attributed to the fact that $\gamma_\eta$ in our simulations varies over a much narrower range and takes significantly lower values ($\gamma_\eta$ = 0.009 – 0.018), compared with the values explored by Wünnemann and Ivanov [2003], where $\gamma_\eta$ = 0.1 and 0.8. We note that the values for $\gamma_\eta$ and $\gamma_\beta$ used in this study are based on previous studies that successfully matched the lunar crater morphology. Some of the differences, however, could stem from the way the vibrational field is calculated in iSALE, as higher velocity impacts will inherently produce a higher vibration velocity field in the near field (close to the point of impact) for the craters of the same size.

We will now explore the linkage between our findings and the observational evidence, and revisit the scaling laws, with the aim to reconcile the differences between the two acoustic fluidization scaling models and shed more light on their applicability.

**6.1 Comparison to Observations**

The onset of transitional craters (e.g., a flat floor but no central peak) is at approximately 13 km for both acoustic fluidization scaling methodologies. The onset of complex structures (e.g., appearance of a central peak) is at 19 km for impactor size scaling and 26 km for the coupling parameter scaling. Thus, the diameters of transitional craters span from 13 – 19 km for impactor size scaling, and 13 – 26 km for the coupling parameter scaling. It is evident that the resulting crater diameters are fairly robust and insensitive to the choice of acoustic fluidization scaling, at least within the scope of our simulations. Therefore, either acoustic fluidization scaling (within the range of $\gamma_\eta$ and $\gamma_\beta$ values tested in this study) is appropriate, if the goal is to study the lateral size and crater growth in the simple-to-complex regime.

However, the preceding statement does not apply to crater morphologies and classification regimes. As evidenced by observations [e.g., *Pike*, 1977b; *Clayton et al.*, 2013], a crater morphology progression (e.g., flat floors, central peak) from simple to complex structures with increasing crater diameter is expected; however, the intricacies of the impactor size/velocity dependence is not well defined in the context of numerical modeling using acoustic fluidization. The onset of complex structures, such as flat floors and central peaks, as a function of the impactor size and velocity combination, appears to be highly sensitive to the acoustic fluidization scaling choice. Using the impactor size scaling implementation considering an approximately constant transient cavity diameter, large/slow impactors will tend to produce early transition from simple to complex crater morphologies, as compared to their small/fast counterparts. In particular, relative to the far ends of the transitional regime spectrum, the largest transitional craters would be those formed by small/fast impactors. The coupling parameter scaling leads to the opposite outcome – small/fast impactors are more likely to lead to the early onset of complex structures. Therefore, the largest transitional craters would be the result of large/slow projectiles. These contradicting outcomes are significant and, as such, should be reconciled.

Since we are interested in evaluating the two acoustic fluidization scaling methods and their applicability for simulating simple-to-complex craters on planetary surfaces, we will briefly touch on experimental data. An extensive discussion is outside the scope of this contribution as there is comprehensive literature available on cratering mechanics and scaling laws [e.g., *Schultz*, 1988; *Holsapple*, 1993; *O'Keefe and Ahrens*, 1993]. Laboratory experiments involving a number of target and projectile materials have revealed that the shape of the crater depends on both the impactor velocity and size [*Schultz and Gault*, 1985a,1985b]. However, it has also been shown that the penetration depth dependency varies according to velocity regimes [*Schultz*, 1988]. For example, impacts at speeds lower than the target sound speed ($c_s$) scale differently than those exceeding $c_s$. This regime dependency might play a role in the lowest impact velocity (6 km/s) in our study, since it is close to $c_s$ of granite. This does not, however, explain the stark contrast between the outcomes of two acoustic fluidization scaling regimes. The issue of energy partition and shock coupling under certain velocity conditions have not been fully resolved, especially when it comes to the implications and direct applicability to specific planetary bodies (e.g., Mars or Mercury). Gault et al. [1975] suggested that for constant kinetic energies and impact velocities exceeding 15 km/s, the fraction of kinetic energy available for heating remains fairly constant. Additional mechanisms, such as energy redistribution due to internal heating [*Cintala and Grieve*, 1984; *Grieve and Cintala*, 1992] and similarity to shallow burst explosions [*Roddy*, 1977], have been proposed as possible reasons for shallow crater depths at high impact velocities. However, we note that our study does not reveal any particular trends in transient cavity *d/D* as a function of impact velocity. A very slight (with slope of ~$10^{-3}$) increase in transient cavity *d/D* for the impactor size scaling is consistent with the lower values of Block-Model acoustic fluidization viscosity constant.

Our simulated craters have depth to diameter ratios consistent with observed lunar craters [*Kalynn et al.*, 2013] (Figure 2b). The coupling parameter scaling produces slightly, but systematically deeper craters, which are broadly in line with the highlands type of surface. The impactor size scaling, however, is even more representative of the observed trends, hinting at being possibly more appropriate mode of acoustic fluidization scaling choice.

The main factor driving the differences in morphology between the two simulation suites is the effect of acoustic fluidization (Figure 8). For example, the simulation using the coupling parameter scaling at 20 km/s is nearly a mirror image, morphologically speaking, of the simulation with the impactor size scaling at 6 km/s, and vice versa. This is especially pronounced in the simulation sets where the transition from simple-to-complex regimes occurs.

As a final step in examining and contrasting the two acoustic fluidization scaling approaches, we compared the modeled craters to the observed lunar craters. Figure 7 shows the results for impact velocities of 6 km/s and 20 km/s for each simulation set, along with lunar crater profiles. Lunar crater profiles represent the craters that have been defined as fresh (dating to the Eratostenian or Copernican era) [*Losiak et al.*, 2009]. All craters, some of which are also shown in Figure 1, are listed in the figure legend. To better estimate whether the simulated craters represent a broad range of the observed lunar craters, the profiles were extracted for both the mare and the highlands, selecting only those craters that are classified as fresh, and which fall within the size regime applicable to the given simulation set.

Figure 7 demonstrates that the lunar craters exhibit a wide range of morphologies even if they have approximately the same diameter. Figure 2b shows the large variability of crater depths on a similar target (mare vs highlands) [cf., *Kalynn et al.*, 2013; *Clayton et al.*, 2013]. While target properties, such as composition, layering and porosity [*Housen and Holsapple*, 2000; *Collins et al.*, 2002; *Wünnemann et al.*, 2006; *Collins et al.*, 2011] can affect temporal evolution and morphology of a crater, they may not account for all of the observed differences, especially for craters of the same diameter and for a similar target. Thus, the transitional (simple-to-complex) regime, while very sensitive to model assumptions, is also an ideal candidate for evaluating the applicability of the acoustic fluidization scaling assumptions, especially for simulations with varying impact velocities.

Across all simulations, the rim height/shape and the crater wall slopes are in good agreement with the observed lunar craters. However, there are slight differences between the impactor size scaling and coupling parameter scaling. In high velocity impacts, the former results in steeper crater walls, whereas the opposite is true for the latter. The rim heights, while comparable between fast and slow impacts in the coupling parameter scaling, are sharper and more pronounced in high velocity impacts using the impactor scaling implementation. This difference increases as a function of crater diameter. For example, there is a notable progression from simulation set 2000 ($D_f$ = 13 km) (Figure 7f) to simulation set 4000 ($D_f$ = 26 km) (Figure 7h). This is not unexpected, as the $T_{dec}$ is longer for large/slow impactors, thereby allowing the crater collapse long after this process seizes in craters produced by small/fast impactors. For large craters ($D_f \geq$ 19 km), the impactor size scaling appears to be more consistent with the observed lunar craters.

Morphologically, the simulated craters are broadly consistent with the observed lunar craters, although the models do not produce terraces and broad flat floors. The simulated craters, however, are too deep, especially when the coupling parameter scaling is applied (Figure 7a-d). This is expected since our models do not include the effect of dilatancy, the creation of porosity in a shearing geological material. Dilatancy is more effective on the Moon than on the Earth, and in craters with $D_f$ < 25 km [*Collins*, 2014]. For a detailed discussion on dilatancy and its implementation in iSALE, the reader is referred to Collins [2014]. It follows that if dilatancy was accounted for, the resulting final craters would be shallower.

Both acoustic fluidization scaling assumptions are relatively simple and may not accurately represent all of the physical details of strength during crater collapse. However, our findings do suggest that the impactor size scaling approach is a better alternative to the coupling parameter scaling, and more appropriate, especially for a sensitive regime such as the simple-to-complex transition.

### 6.2 Relation to the Late-Stage Equivalence Principle and Implications for Other Planetary Bodies

In this study, the point source coupling parameter was adopted for the purpose of 'bookkeeping', used as a starting point to predict the transient cavity size and calculate the impactor size and velocity combination. Indeed, as evident from our results, the coupling parameter is undoubtedly a convenient and useful relation, as it correctly predicts the transient cavity diameters and, somewhat, the transient cavity depths. It is not suitable, however, for predicting the crater morphology, final crater depths or temporal evolution beyond the transient cavity stage, as evidenced by the divergent outcomes in the transitional crater regime.

While our aim was not to explicitly test the scope of the coupling parameter [*Holsapple and Schmidt*, 1987] and, by extension, the late-stage equivalence principle [*Dienes and Walsh*, 1970], it is important to comment on its practicality, as it pertains to the specific problem presented in our study. In principle, if the late-stage equivalence holds, then it is expected that the exact same shock and rarefaction will produce exactly the same fragment sizes and, therefore, acoustic energy field. In that case, the coupling parameter scaling would be more appropriate.

However, the size of the region at which the point source assumption holds valid has been greatly debated. It has been suggested that this approximation should hold as long at the phenomena investigated lies beyond approximately one impactor radius [*Housen and Holsapple*, 2011] or several impactor radii [*Pierazzo et al.*, 1997]. Note the transient cavity diameters are 7-16 times larger than our projectile sizes, indicating that much of the collapsing material is from the near field, where late-stage equivalence may not apply. Furthermore, Housen and Holsapple [2011] argue that the shock and rarefaction are not necessarily identical, but simply result in the same size transient cavity. This is consistent with our findings. Indeed, our results demonstrate that the transitional crater regime is highly susceptible to the intricate interplay between the impactor size and velocity.

Another consideration, briefly explored and tested by Kenkmann et al. [2006], but otherwise largely neglected, is that the block size (and consequently the viscosity of the fluidized target) may depend on distance to the point of impact. The small amount of observational data, however, hinders efforts to perform systematic studies and constrain the models. The Block-Model of acoustic fluidization assumes constant block size, which may be a rough approximation.

Recent discrete element modeling of long runout landslides suggests acoustic fluidization is responsible for these large landslides extraordinary mobility [*Johnson et al.*, 2016b]. This modeling suggests that the frequency of acoustic vibrations driving fluidization is set by the size of grains making up the sliding mass [*Johnson et al.*, 2016b]. This supports some of the underlying assumptions of the Block-Model of acoustic fluidization. With the recent addition of fragmentation calculation in iSALE [*Johnson et al.*, 2016c] to self consistently calculate the appropriate block size without the need for the empirical parameters $\gamma_\eta$ and $\gamma_\beta$. Such a calculation would allow one to explore the effect of impact velocity directly and may be the subject of future work.

Mars and Mercury, despite having very similar gravity (3.72 m/s$^2$ and 3.70 m/s$^2$, respectively), exhibit notable difference in crater diameters at which the simple-to-complex transition occurs. On Mercury, the onset of complex structures (e.g., flat floor, terracing) occurs at a factor of 1.5-2 larger than expected diameters [*Garvin and Frawley*, 1998; *Barnouin et al.*, 2011; *Susorney et al.*, 2016], even when gravity is taken into consideration [*Pike*, 1980a; Pike, 1988; *Barnouin et al.*, 2011]. This has been attributed to high median impact velocity on Mercury (42.5 km/s) relative to Mars [*Lefeuvre and Wieczorek*, 2008]. The actual impact velocity range is from ~18 – 135 km/s [*Minton and Malhotra*, 2010]. For comparison, the RMS impact velocity on Mars is only about 13 km/s [e.g., *Hartmann*, 1981; *Lefeuvre and Wieczorek*, 2008]. Our study suggests that in order to account for the observed trend of later onset of complex structures on Mercury, the scaling by impactor size would be more appropriate. Additional studies employing high impact velocities, consistent with those at Mercury, are recommended to further examine our assertion. While our study did not investigate the overall effect of the target, future work should explore the link between the varying target properties (e.g., porosity, layering, etc.), impact velocity and progression in crater morphology, as well as trends for a wider range of crater sizes (outside of the transitional regime).

## 7. Conclusions

Impact cratering is a geological process controlled by many factors. Both the properties of the target and the impactor are important and to better understand the effect of each, these often need to be decoupled. Nevertheless, the impact velocity does play a role, albeit secondary compared to target and gravity, in the impact excavation stage [e.g., *Xiao et al.*, 2014] and, thus, should not be neglected. In this study, we investigate the role of impact velocity on crater formation under the assumption that acoustic fluidization is the physical explanation for the transition from simple to complex craters. We employ two commonly adopted acoustic fluidization Block-Size scaling methodologies, scaling by coupling parameter [*Ivanov and Artemieva*, 2002] and scaling by impactor size [*Wünnemann and Ivanov*, 2003]. To best describe the underlying assumptions and analyze the outcomes, we varied the impactor size and velocity, while keeping all other parameters constant. The conclusions of our study are as follows:

i. The effect of acoustic fluidization model parameters on simple-to-complex lunar craters can be significant, and a special consideration should be given when choosing an appropriate Block-Model acoustic fluidization scaling methodology.

ii. While lateral crater growth is relatively insensitive to the choice of the acoustic fluidization scaling, within the range of model parameters investigated in this study, the crater depth and morphology, especially in the transitional crater regime are highly sensitive to the choice of acoustic fluidization model parameters and how they scale with impactor size.

iii. The crater size regime at which the effect of scaling choice is most relevant is the simple-to-complex regime, specifically in the final crater diameter range from 13–20 km as applicable to lunar craters. In the complex crater regime, the differences are non-negligible, and are reflected in the crater depth and morphology of the central peak. Simple crater formation ($D_f$ = 10 km), however, is relatively insensitive to the choice of the acoustic fluidization scaling (coupling parameter or impactor size).

iv. The scaling by impactor size is appropriate for varying impactor sizes across invariant impact velocities. Impactor size scaling also appears to be consistent with the observations of crater simple-to-complex transition on Mercury and is, hence, applicable for modeling simple-to-complex craters on planetary surfaces. However, more comparative studies are needed to verify our assertion.

v. Our study also suggests that the effect of impact velocity on the transition from simple to complex structures on solid planetary surfaces is more significant than previously thought.

**Acknowledgements**

The authors thank G. Collins and K. Wünnemann for their helpful comments to improve this paper, and T. Bowling for helpful comments on the earlier version of the manuscript. EAS gratefully acknowledges the Natural Sciences and Engineering Research Council of Canada Postdoctoral Fellowship program for partly funding this project. This work was funded by the Natural Sciences and Engineering Research Council of Canada through various grants to GRO. We gratefully acknowledge the developers of iSALE2D (www.isale-code.de), the simulation code used in our research, including G. Collins, K. Wünnermann, B. Ivanov, D. Elbeshausen and

J. Melosh. All data associated with this study are available in tables. The simulation outputs are available upon request from elizabeth_silber@brown.edu.

## References

Amsden, A., H. Ruppel, and C. Hirt (1980), SALE: A simplified ALE computer program for fluid flow at all speeds*Rep.*, 101 pp, LANL, Los Alamos, New Mexico.

Baker, D. M. H., J. W. Head, G. S. Collins, and R. W. K. Potter (2016), The formation of peak-ring basins: Working hypotheses and path forward in using observations to constrain models of impact-basin formation, *Icarus*, *273*, 146-163, doi:10.1016/j.icarus.2015.11.033.

Barnouin, O., C. Ernst, J. Heinick, S. Sugita, M. Cintala, D. Crawford, and T. Matsui (2011), Experimental results investigating the impact velocity effects on crater growth and the transient crater diameter-to-depth ratio, paper presented at Lunar and Planetary Science Conference.

Benz, W., A. G. W. Cameron, and H. J. Melosh (1989), The origin of the Moon and the single-impact hypothesis III, *Icarus*, *81*(1), 113-131, doi:10.1016/0019-1035(89)90129-2.

Billingsley, J. (1969), Comparison of experimental and predicted axial pressure variation for semi-infinite metallic targets, in *Hypervelocity Impact Conference*, edited, American Institute of Aeronautics and Astronautics, doi:10.2514/6.1969-361.

Bray, V. J., and P. M. Schenk (2015), Pristine impact crater morphology on Pluto – Expectations for New Horizons, *Icarus*, *246*, 156-164, doi:10.1016/j.icarus.2014.05.005.

Cintala, M., and R. Grieve (1984), Energy partitioning during terrestrial impact events: Melt production and scaling laws, paper presented at Lunar and Planetary Science Conference.

Clayton, J., G. R. Osinski, L. L. Tornabene, J. Kalynn, and C. L. Johnson (2013), paper presented at 44th Lunar and Planetary Science Conference.

Collins, G., H. J. Melosh, J. V. Morgan, and M. R. Warner (2002), Hydrocode Simulations of Chicxulub Crater Collapse and Peak-Ring Formation, *Icarus*, *157*(1), 24-33, doi:10.1006/icar.2002.6822.

Collins, G. S. (2014), Numerical simulations of impact crater formation with dilatancy, *Journal of Geophysical Research: Planets*, *119*(12), 2600-2619, doi:10.1002/2014je004708.

Collins, G. S., H. J. Melosh, and B. A. Ivanov (2004), Modeling damage and deformation in impact simulations, *Meteoritics & Planetary Science*, *39*(2), 217-231, doi:10.1111/j.1945-5100.2004.tb00337.x.

Collins, G. S., H. J. Melosh, and K. Wünnemann (2011), Improvements to the ε-α porous compaction model for simulating impacts into high-porosity solar system objects, *International Journal of Impact Engineering*, *38*(6), 434-439, doi:10.1016/j.ijimpeng.2010.10.013.

Collins, G. S., J. Morgan, P. Barton, G. L. Christeson, S. Gulick, J. Urrutia, M. Warner, and K. Wünnemann (2008), Dynamic modeling suggests terrace zone asymmetry in the Chicxulub crater is caused by target heterogeneity, *Earth and Planetary Science Letters*, *270*(3-4), 221-230, doi:10.1016/j.epsl.2008.03.032.

Collins, G. S., and K. Wünnemann (2005), How big was the Chesapeake Bay impact? Insight from numerical modeling, *Geology*, *33*(12), 925, doi:10.1130/g21854.1.

Crawford, D. A., and P. Schultz (2013), A model of localized shear heating with implications for the morphology and paleomagnetism of complex craters, paper presented at LPI Contributions.

Davison, T. M., G. S. Collins, and F. J. Ciesla (2010), Numerical modelling of heating in porous planetesimal collisions, *Icarus*, *208*(1), 468-481, doi:10.1016/j.icarus.2010.01.034.

Dence, M., R. A. Grieve, and P. Robertson (1977), Terrestrial impact structures-Principal characteristics and energy considerations, paper presented at Impact and explosion cratering: Planetary and terrestrial implications, September 13-17, 1976.

Dienes, J., and J. Walsh (1970), Theory of impact: Some general principles and the method of Eulerian codes, *High-velocity impact phenomena*, *45*.

Elbeshausen, D., K. Wünnemann, and G. S. Collins (2009), Scaling of oblique impacts in frictional targets: Implications for crater size and formation mechanisms, *Icarus*, *204*(2), 716-731, doi:10.1016/j.icarus.2009.07.018.

Garvin, J. B., and J. J. Frawley (1998), Geometric properties of Martian impact craters: Preliminary results from the Mars Orbiter Laser Altimeter, *Geophysical Research Letters*, *25*(24), 4405-4408, doi:10.1029/1998gl900177.

Gault, D. E., J. E. Guest, J. B. Murray, D. Dzurisin, and M. C. Malin (1975), Some comparisons of impact craters on Mercury and the Moon, *Journal of Geophysical Research*, *80*(17), 2444-2460, doi:10.1029/JB080i017p02444.

Grieve, R. A. F., and M. J. Cintala (1992), An analysis of differential impact melt-crater scaling and implications for the terrestrial impact record, *Meteoritics*, *27*(5), 526-538, doi:10.1111/j.1945-5100.1992.tb01074.x.

Grieve, R. A. F., and A. M. Therriault (2004), Observations at terrestrial impact structures: Their utility in constraining crater formation, *Meteoritics & Planetary Science*, *39*(2), 199-216, doi:10.1111/j.1945-5100.2004.tb00336.x.

Hartmann, W. (1981), Discovery of multi-ring basins-Gestalt perception in planetary science, paper presented at Multi-ring basins: Formation and Evolution.

Hay, H. C. F. C., G. S. Collins, and T. M. Davison (2014), Complex crater collapse: a comparison of the block and Melosh models of acoustic fluidization, paper presented at 45th Lunar and Planetary Science Conference.

Holsapple, K. (1983), On the existence and implications of coupling parameters in cratering mechanics, paper presented at Lunar and Planetary Science Conference.

Holsapple, K. A. (1981), Coupling parameters in cratering, edited, p. 494, Eos Trans. AGU.

Holsapple, K. A. (1993), The scaling of impact processes in planetary sciences, *Annual Review of Earth and Planetary Sciences*, *21*, 333-373, doi:10.1146/annurev.ea.21.050193.002001.

Holsapple, K. A., and R. M. Schmidt (1982), On the scaling of crater dimensions: 2. Impact processes, *Journal of Geophysical Research*, *87*(B3), 1849, doi:10.1029/JB087iB03p01849.

Holsapple, K. A., and R. M. Schmidt (1987), Point source solutions and coupling parameters in cratering mechanics, *Journal of Geophysical Research*, *92*(B7), 6350, doi:10.1029/JB092iB07p06350.

Housen, K. R., and K. A. Holsapple (2000), Numerical Simulations of Impact Cratering in Porous Materials, paper presented at 31st Annual Lunar and Planetary Science Conference, March 13-17, 2000.

Housen, K. R., and K. A. Holsapple (2011), Ejecta from impact craters, *Icarus*, *211*(1), 856-875, doi:10.1016/j.icarus.2010.09.017.

Housen, K. R., R. M. Schmidt, and K. A. Holsapple (1983), Crater ejecta scaling laws: Fundamental forms based on dimensional analysis, *Journal of Geophysical Research*, *88*(B3), 2485, doi:10.1029/JB088iB03p02485.

Huang, S. L., and P. C. Chou (1968), Calculations of Expanding Shock Waves and Late-Stage Equivalence*Rep.*, DTIC Document.

Ivanov, B., D. Deniem, and G. Neukum (1997), Implementation of dynamic strength models into 2D hydrocodes: Applications for atmospheric breakup and impact cratering, *International Journal of Impact Engineering*, *20*(1), 411-430.

Ivanov, B., G. Kocharyan, V. Kostuchenko, A. Kirjakov, and L. Pevzner (1996), Puchezh-Katunki impact crater: Preliminary data on recovered core block structure, paper presented at Lunar and Planetary Science.

Ivanov, B. A., and N. A. Artemieva (2002), Numerical modeling of the formation of large impact craters, *Geological Society of America Special Papers*, *356*, 619-630, doi:10.1130/0-8137-2356-6.619.

Ivanov, B. A., and E. P. Turtle (2001), Modeling impact crater collapse: acoustic fluidization implemented into a hydrocode. Paper presented at 32nd Lunar and Planetary Science Conference.

Ivanov, B. A., and V. N. Kostuchenko (1997), Block oscillation model for impact crater collapse, paper presented at Lunar and Planetary Science Conference XXVIII.

Ivanov, B. A., H. J. Melosh, and E. Pierazzo (2010), Basin-forming impacts: Reconnaissance modeling, Geol. Soc. Am. Spec. Papers, 465, 29–49, doi:10.1130/2010.2465(03).

Jaeger, J. C., N. G. Cook, and R. Zimmerman (2009), *Fundamentals of rock mechanics*, John Wiley & Sons.

Johnson, B. C., G. S. Collins, D. A. Minton, T. J. Bowling, B. M. Simonson, and M. T. Zuber (2016a), Spherule layers, crater scaling laws, and the population of ancient terrestrial impactors, *Icarus*, *271*, 350-359, doi:10.1016/j.icarus.2016.02.023.

Johnson, B. C., C. S. Campbell, and H. J. Melosh (2016b), The reduction of friction in long runout landslides as an emergent phenomenon, *Journal of Geophysical Research: Earth Surface*, *121*(5), 881-889, doi:10.1002/2015JF003751.

Johnson, B., T. Bowling, and H. Melosh (2016c), Steps Toward Implementing the Grady-Kipp Fragmentation Model in an Eulerian Hydrocode, paper presented at Lunar and Planetary Science Conference.

Kalynn, J., C. L. Johnson, G. R. Osinski, and O. Barnouin (2013), Topographic characterization of lunar complex craters, *Geophysical Research Letters*, *40*(1), 38-42, doi:10.1029/2012gl053608.

Kenkmann, T., A. Jahn, and K. Wünnemann (2006), " Block Size" in a Complex Impact Crater Inferred from the Upheaval Dome Structure, Utah, (abstract #1540), paper presented at 37th Annual Lunar and Planetary Science Conference.

Kenkmann, T., Reimold, W.U., Khirfan, M. et al. (2010), The complex impact crater Jebel Waqf as Suwwan in Jordan: effects of target heterogeneity and impact obliquity on central uplift formation, in Large Meteorite Impacts and Planetary Evolution IV (eds R.L. Gibson and W.U. Reimold), Geological Society of America Special Paper 465, Geological Society of America, Boulder, CO, pp. 471–487

Kenkmann, T., Collins, G.S. and K. Wünnemann, (2012), The modification stage of crater formation. Impact Cratering: Processes and Products, pp.60-75.

Le Feuvre, M., and M. A. Wieczorek (2011), Nonuniform cratering of the Moon and a revised crater chronology of the inner Solar System, *Icarus*, *214*(1), 1-20, doi:10.1016/j.icarus.2011.03.010.

Lefeuvre, M., and M. Wieczorek (2008), Nonuniform cratering of the terrestrial planets, *Icarus*, *197*(1), 291-306, doi:10.1016/j.icarus.2008.04.011.

Losiak, A., D. Wilhelms, C. Byrne, K. Thaisen, S. Weider, T. Kohout, K. O'Sullivan, and D. Kring (2009), A new lunar impact crater database, paper presented at Lunar and Planetary Science Conference.

McKinnon, W. B. (1978), An Investigation Into the Role of Plastic Failure in Crater Modification.

Melosh, H. J. (1977), Crater modification by gravity.

Melosh, H. J. (1979), Acoustic fluidization: A new geologic process?, *Journal of Geophysical Research: Solid Earth*, *84*(B13), 7513-7520, doi:10.1029/JB084iB13p07513.

Melosh, H. J. (1984), Impact ejection, spallation, and the origin of meteorites, *Icarus*, *59*(2), 234-260, doi:10.1016/0019-1035(84)90026-5.

Melosh, H. J. (1989), Impact cratering: A geologic process, *Research supported by NASA. New York, Oxford University Press (Oxford Monographs on Geology and Geophysics, No. 11), 1989, 253 p.*, *1*.

Melosh, H. J., et al. (2013), The origin of lunar mascon basins, *Science*, *340*(6140), 1552-1555, doi:10.1126/science.1235768.

Melosh, H. J., and A. Ivanov (1999), impact crater collapse.

Melosh, H. J., E. V. Ryan, and E. Asphaug (1992), Dynamic fragmentation in impacts: Hydrocode simulation of laboratory impacts, *Journal of Geophysical Research: Planets*, *97*(E9), 14735-14759, doi:10.1029/92JE01632.

Milbury, C., B. C. Johnson, H. J. Melosh, G. S. Collins, D. M. Blair, J. M. Soderblom, F. Nimmo, C. J. Bierson, R. J. Phillips, and M. T. Zuber (2015), Preimpact porosity controls the gravity signature of lunar craters, *Geophysical Research Letters*, *42*(22), 9711-9716, doi:10.1002/2015GL066198.

Miljković, K., G. S. Collins, S. Mannick, and P. A. Bland (2013), Morphology and population of binary asteroid impact craters, *Earth and Planetary Science Letters*, *363*, 121-132, doi:10.1016/j.epsl.2012.12.033.

Minton, D. A., and R. Malhotra (2010), Dynamical erosion of the asteroid belt and implications for large impacts in the inner Solar System, *Icarus*, *207*(2), 744-757, doi:10.1016/j.icarus.2009.12.008.

O'Keefe, J. D., and T. J. Ahrens (1993), Planetary cratering mechanics, *Journal of Geophysical Research: Planets*, *98*(E9), 17011-17028, doi:10.1029/93JE01330.

Osinski, G. R., and E. Pierazzo (2012), *Impact cratering: Processes and products*, John Wiley & Sons.

Pearce, S. J., and H. J. Melosh (1986), Terrace width variations in complex lunar craters, *Geophysical Research Letters*, *13*(13), 1419-1422, doi:10.1029/GL013i013p01419.

Pierazzo, E., et al. (2008), Validation of numerical codes for impact and explosion cratering: Impacts on strengthless and metal targets, *Meteoritics & Planetary Science*, *43*(12), 1917-1938, doi:10.1111/j.1945-5100.2008.tb00653.x.

Pierazzo, E., D. A. Kring, and H. J. Melosh (1998), Hydrocode simulation of the Chicxulub impact event and the production of climatically active gases, *Journal of Geophysical Research: Planets*, *103*(E12), 28607-28625, doi:10.1029/98je02496.

Pierazzo, E., A. M. Vickery, and H. J. Melosh (1997), A Reevaluation of Impact Melt Production, *Icarus*, *127*(2), 408-423, doi:10.1006/icar.1997.5713.

Pike, R. J. (1977a), Apparent depth-apparent diameter relation for lunar craters, paper presented at 8th Lunar Science Conference, Houston, Texas, USA, March 14-18, 1977.

Pike, R. J. (1977b), Size-dependence in the shape of fresh impact craters on the moon, paper presented at The Symposium on Planetary Cratering Mechanics, Flagstaff, Arizona, USA, September 13-17, 1976.

Pike, R. J. (1980a), Control of crater morphology by gravity and target type-Mars, Earth, Moon, paper presented at Lunar and Planetary Science Conference Proceedings.

Pike, R. J. (1980b), Control of crater morphology by gravity and target type - Mars, Earth, Moon, paper presented at Lunar and Planetary Science Conference 11th Proceedings, New York, Pergamon Press, Houston, TX, March 17-21, 1980.

Potter, R. W. K., G. S. Collins, W. S. Kiefer, P. J. McGovern, and D. A. Kring (2012), Constraining the size of the South Pole-Aitken basin impact, *Icarus*, *220*(2), 730-743, doi:10.1016/j.icarus.2012.05.032.

Rae, A. S. P., Collins, G. S., Grieve, R. A. F., Osinski, G. R., and J. V. Morgan (2017), Complex crater formation: Insights from combining observations of shock pressure distribution with numerical models at the West Clearwater Lake impact structure. *Meteoritics & Planetary Science*, 1-21, doi: 10.1111/maps.12825

Roddy, D. J. (1977), Large-scale impact and explosion craters.

Sakurai, A. (1964), Blast wave theory *Rep.*, DTIC Document, USA.

Schmidt, R. M., and K. A. Holsapple (1982), Estimates of crater size for large-body impact: Gravity-scaling results, *Geological Society of America Special Papers*, *190*, 93-102, doi:10.1130/SPE190-p93.

Schmidt, R. M., and K. R. Housen (1987), Some recent advances in the scaling of impact and explosion cratering, *International Journal of Impact Engineering*, *5*(1), 543-560.

Schultz, P., and D. Gault (1985a), The effect of projectile shape on cratering efficiency and crater profile in granular targets, paper presented at Lunar and Planetary Science Conference.

Schultz, P. H. (1988), Cratering on Mercury: A relook, *Mercury*, 274-335.

Schultz, P. H., and D. E. Gault (1985b), Clustered impacts: Experiments and implications, *Journal of Geophysical Research: Solid Earth*, *90*(B5), 3701-3732, doi:10.1029/JB090iB05p03701.

Scott, R. G., and K. Benn (2001), Peak-ring rim collapse accommodated by impact melt-filled transfer faults, Sudbury impact structure, Canada, *Geology*, *29*(8), 747-750, doi:10.1130/0091-7613(2001)029.

Sedov, L. I. (1959), *Similarity and dimensional methods in mechanics*, Academic Press, New York.

Senft, L. E., and S. T. Stewart (2009), Dynamic fault weakening and the formation of large impact craters, *Earth and Planetary Science Letters*, *287*(3-4), 471-482, doi:10.1016/j.epsl.2009.08.033.

Spray, J. G., and L. M. Thompson (1995), Friction melt distribution in a multi-ring impact basin, *Nature*, *373*(6510), 130-132.

Susorney, H. C. M., O. S. Barnouin, C. M. Ernst, and C. L. Johnson (2016), Morphometry of impact craters on Mercury from MESSENGER altimetry and imaging, *Icarus*, *271*, 180-193, doi:10.1016/j.icarus.2016.01.022.

Svetsov, V. V., and V. V. Shuvalov (2015), Water delivery to the Moon by asteroidal and cometary impacts, *Planetary and Space Science*, *117*, 444-452, doi:10.1016/j.pss.2015.09.011.

Taylor, G. (1950), The Formation of a Blast Wave by a Very Intense Explosion. I. Theoretical Discussion, *Proceedings of the Royal Society of London. Series A, Mathematical and Physical Sciences*, *201*(1065), 159-174.

Wünnemann, K., G. S. Collins, and H. J. Melosh (2006), A strain-based porosity model for use in hydrocode simulations of impacts and implications for transient crater growth in porous targets, *Icarus*, *180*(2), 514-527, doi:10.1016/j.icarus.2005.10.013.

Wünnemann, K., G. S. Collins, and G. R. Osinski (2008), Numerical modelling of impact melt production in porous rocks, *Earth and Planetary Science Letters*, *269*(3-4), 530-539, doi:10.1016/j.epsl.2008.03.007.

Wünnemann, K. (2001), Die numerische Behandlung von Impaktprozessen - Kraterbildung, stosswelleninduzierte Krustenmodifikationen und ozeanische Einschlagereignisse, PhD thesis, Westfaelische Wilhelms-Universitaet, Münster, 181 p.

Wünnemann, K., and B. A. Ivanov (2003), Numerical modelling of the impact crater depth–diameter dependence in an acoustically fluidized target, *Planetary and Space Science*, *51*(13), 831-845, doi:10.1016/j.pss.2003.08.001.

Xiao, Z., R. G. Strom, C. R. Chapman, J. W. Head, C. Klimczak, L. R. Ostrach, J. Helbert, and P. D'Incecco (2014), Comparisons of fresh complex impact craters on Mercury and the Moon: Implications for controlling factors in impact excavation processes, *Icarus*, *228*, 260-275, doi:10.1016/j.icarus.2013.10.002.

Yue, Z., B. C. Johnson, D. A. Minton, H. J. Melosh, K. Di, W. Hu, and Y. Liu (2013), Projectile remnants in central peaks of lunar impact craters, *Nature Geoscience*, *6*(6), 435-437, doi:10.1038/ngeo1828.

**Tables and Table Captions:**

**Table 1:** The input parameters for simulations.

| Description | Impactor | Target | Reference |
|---|---|---|---|
| Poisson ratio | 0.25 | 0.25 | a, b |
| Melt temperature at zero pressure (K) | 1373 | 1673 | b, c |
| Thermal softening parameter | 1.1 | 1.2 | c, d |
| Constant in Simon approximation, *a* (GPa) | 1.4 | 6.0 | a, b |
| Exponent in Simon approximation, *c* | 4.05 | 3.00 | a, b, c |
| Cohesion, intact (MPa) | 10 | 10 | b, e |
| Coefficient of internal friction, intact | 1.2 | 2.0 | b, e |
| Limiting strength at high pressure, intact (GPa) | 3.5 | 2.5 | b, e |
| Cohesion, damaged (MPa) | 0.01 | 0.01 | b, e |
| Coefficient of internal friction, damaged | 0.6 | 0.7 | b, c, e |
| Limiting strength at high pressure, damaged (GPa) | 3.5 | 2.5 | b, e |
| Equation of state (EOS) (ANEOS) | Dunite | Granite | f, g |

[a] Miljkovic et al. (2013)
[b] Yue et al. (2013)
[c] Davison et al. (2010)
[d] Potter et al. (2012)
[e] Collins et al. (2004)
[f] Bentz et al. (1998)
[g] Pierazzo et al. (1997)

**Table 2:** The simulation results, sorted by the acoustic fluidization scaling, as shown in the left side bar. The upper and the lower regions of the Table represent the simulations using the coupling parameter/transient diameter scaling, and scaling by the impactor size, respectively. Each column variable is described on the bottom of the Table. The crater type notation also includes mixed crater types, such as ST (simple-to-transitional, with no clear delineation) and TC (transitional-to-complex, with no clear delineation). Minor differences in the impactor diameter are due to the choice of $\mu$ ($\mu$ = 0.55 for all simulations, except in simulation sets 3000 and 4000, where $\mu$ = 0.56), although this had no influence on the final outcomes, as it can be seen from the table. Due to prohibitively long time it would take to perform simulations for the smallest and fastest impactors, simulations 103 and 1030 were stopped before reaching their intended completion time.

| | Sim # | $D_i$ (m) | $v_i$ (km/s) | $\nu_{lim}$ ($10^4$)($m^2/s$) | $T_{dec}$ (s) | $E_k$ ($10^{19}$)(J) | $V_t$ ($km^3$) | $D_t$ (km) | $d_t$ (km) | $D_f$ (km) | $d_{f(max)}$ (km) | Time (s) | $\pi_2$ ($10^{-6}$) | $\pi_D$ | Crater type |
|---|---|---|---|---|---|---|---|---|---|---|---|---|---|---|---|
| Coupling Parameter Scaling | 100 | 809 | 6 | 1.81 | 14.5 | 1.65 | 56.3 | 6.9 | 2.71 | 9.9 | 2.70 | 315 | 60.4 | 9.81 | S |
| | 101 | 607 | 10 | 1.81 | 14.5 | 1.94 | 53.9 | 6.7 | 2.70 | 9.9 | 2.56 | 315 | 16.3 | 12.70 | S |
| | 102 | 484 | 15 | 1.81 | 14.5 | 2.21 | 51.5 | 6.7 | 2.69 | 9.8 | 2.45 | 315 | 5.8 | 15.90 | S |
| | 103 | 412 | 20 | 1.81 | 14.5 | 2.43 | 49.2 | 6.5 | 2.64 | 9.6 | 2.52 | 305 | 2.8 | 18.20 | S |
| | 200 | 1153 | 6 | 2.58 | 20.7 | 4.78 | 136.9 | 9.3 | 3.52 | 13.7 | 3.23 | 515 | 86.1 | 9.24 | S |
| | 201 | 866 | 10 | 2.58 | 20.7 | 5.63 | 130.6 | 8.9 | 3.55 | 13.4 | 3.16 | 515 | 23.3 | 11.78 | S |
| | 202 | 690 | 15 | 2.58 | 20.7 | 6.41 | 125.5 | 8.9 | 3.35 | 13.3 | 3.02 | 515 | 8.2 | 14.88 | ST |
| | 203 | 587 | 20 | 2.58 | 20.7 | 7.03 | 120.5 | 8.8 | 3.32 | 13.1 | 3.23 | 515 | 3.9 | 17.18 | ST |
| | 300 | 1791 | 6 | 4.01 | 32.2 | 17.94 | 395.7 | 13.5 | 4.84 | 19.7 | 4.37 | 550 | 133.8 | 8.67 | T |
| | 301 | 1345 | 10 | 4.01 | 32.2 | 21.12 | 391.4 | 13.3 | 4.84 | 19.8 | 3.73 | 550 | 36.2 | 11.31 | T |
| | 302 | 1072 | 15 | 4.01 | 32.2 | 24.05 | 373.9 | 12.8 | 4.93 | 19.8 | 3.51 | 550 | 12.8 | 13.71 | T |
| | 303 | 912 | 20 | 4.01 | 32.2 | 26.37 | 359.9 | 12.7 | 4.56 | 19.8 | 3.50 | 550 | 6.1 | 16.03 | T |
| | 400 | 2446 | 6 | 5.48 | 43.9 | 45.69 | 840.2 | 17.2 | 6.24 | 25.6 | 3.91 | 755 | 182.7 | 8.10 | T |
| | 401 | 1837 | 10 | 5.48 | 43.9 | 53.80 | 840.5 | 17.0 | 6.34 | 26.7 | 3.51 | 755 | 49.4 | 10.62 | C |
| | 402 | 1464 | 15 | 5.48 | 43.9 | 61.25 | 806.1 | 16.8 | 6.15 | 25.9 | 3.30 | 755 | 17.5 | 13.15 | C |
| | 403 | 1246 | 20 | 5.48 | 43.9 | 67.16 | 776.3 | 16.6 | 5.89 | 25.7 | 3.40 | 755 | 8.4 | 15.34 | C |
| Impactor Size Scaling | 1000 | 809 | 6 | 3.03 | 24.3 | 1.65 | 50.0 | 6.7 | 2.51 | 9.9 | 2.44 | 315 | 60.4 | 9.47 | S |
| | 1010 | 607 | 10 | 2.28 | 18.2 | 1.95 | 51.8 | 6.7 | 2.64 | 9.9 | 2.57 | 315 | 16.4 | 12.69 | S |
| | 1020 | 484 | 15 | 1.81 | 14.5 | 2.21 | 51.5 | 6.7 | 2.69 | 9.8 | 2.45 | 315 | 5.8 | 15.90 | S |
| | 1030 | 412 | 20 | 1.55 | 12.4 | 2.43 | 50.4 | 6.6 | 2.64 | 9.6 | 2.61 | 300 | 2.8 | 18.45 | S |
| | 2000 | 1153 | 6 | 4.32 | 34.6 | 4.78 | 118.9 | 9.0 | 3.34 | 13.1 | 2.54 | 515 | 86.1 | 9.01 | T |
| | 2010 | 866 | 10 | 3.25 | 26.0 | 5.62 | 125.2 | 9.0 | 3.38 | 13.6 | 2.94 | 515 | 23.3 | 12.00 | T |
| | 2020 | 690 | 15 | 2.58 | 20.7 | 6.41 | 125.8 | 9.1 | 3.43 | 13.3 | 3.02 | 515 | 8.2 | 14.88 | T |
| | 2030 | 587 | 20 | 2.20 | 17.6 | 7.02 | 122.5 | 8.8 | 3.46 | 13.0 | 3.06 | 515 | 4.0 | 17.18 | ST |
| | 3000 | 1675 | 6 | 6.28 | 50.3 | 14.70 | 302.1 | 12.5 | 4.44 | 19.0 | 2.44 | 550 | 125.0 | 8.56 | C |
| | 3010 | 1300 | 10 | 4.88 | 39.0 | 19.10 | 340.1 | 12.7 | 4.55 | 19.1 | 2.99 | 550 | 35.0 | 11.20 | TC |
| | 3020 | 1072 | 15 | 4.01 | 32.2 | 24.05 | 373.9 | 12.8 | 4.93 | 19.8 | 3.51 | 550 | 12.8 | 13.71 | T |
| | 3030 | 912 | 20 | 3.42 | 27.4 | 26.30 | 368.8 | 12.7 | 4.79 | 19.0 | 3.91 | 550 | 6.1 | 16.02 | T |
| | 4000 | 2370 | 6 | 8.89 | 71.1 | 41.60 | 698.5 | 16.5 | 6.04 | 26.2 | 2.26 | 755 | 177.0 | 7.98 | C |
| | 4010 | 1838 | 10 | 6.89 | 55.1 | 53.90 | 795.6 | 16.6 | 6.34 | 26.4 | 2.85 | 755 | 49.4 | 10.39 | C |
| | 4020 | 1500 | 15 | 5.63 | 45.0 | 65.90 | 849.3 | 17.0 | 6.38 | 26.4 | 3.47 | 755 | 17.9 | 13.04 | C |
| | 4030 | 1300 | 20 | 4.88 | 39.0 | 76.20 | 880.4 | 17.0 | 6.37 | 26.4 | 3.83 | 755 | 8.7 | 14.99 | TC |

$D_i$ - impactor diameter
$v_i$ - impactor velocity
$\nu_{lim}$ - limiting viscosity (AF)
$T_{dec}$ - decay time (AF)
$E_k$ - impact energy
$V_t$ - transient crater volume
$D_t$ - transient crater diameter
$d_t$ - transient crater depth
$D_f$ - final crater diameter
$d_f$ - final crater depth
S - simple
T - transitional
C - complex

**Figures and Figure Captions:**

**Figure 1:** The progression from simple to complex craters on the Moon: simple (a,b), transitional (c-e) and complex (f) craters. The craters, with their final diameters ($D_f$) are: (a) Kepler A ($D_f$ = 11 km), (b) Cayley ($D_f$ = 14 km), (c) Marius A ($D_f$ = 15 km), (d) Bessel ($D_f$ = 15 km), (e) Picard ($D_f$ = 22 km) and (f) Euler ($D_f$ = 27 km). The images were obtained from the Lunar Reconnaissance Orbiter (LRO) topographic map database (http://target.lroc.asu.edu/q3/).

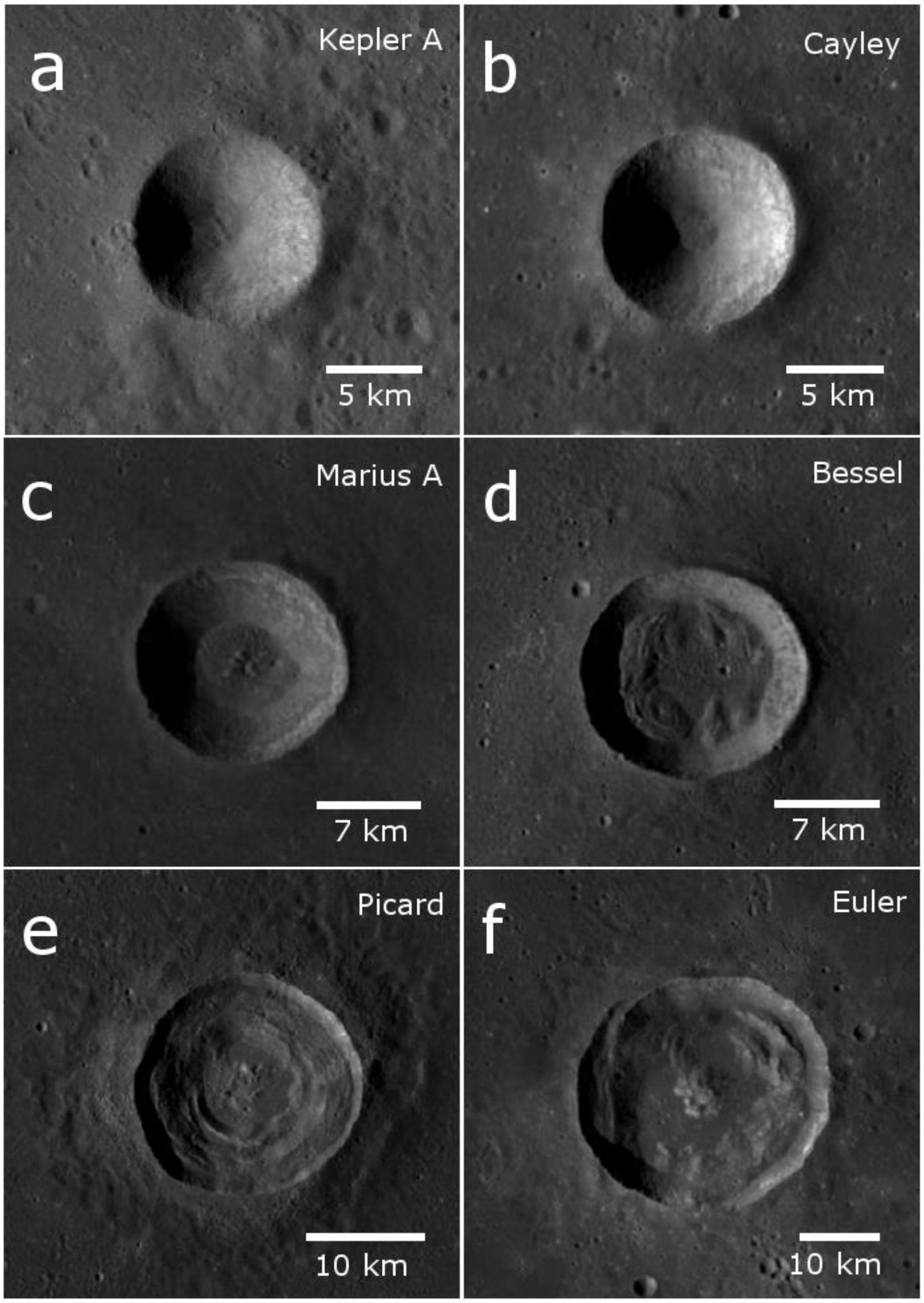

**Figure 2:** (a) Transient crater depth versus diameter. (b) The final crater depth versus diameter plotted against the observed points for transitional to complex craters on the Moon [Kalynn et al., 2013]. The letter notation for the plotted craters is as follows: S – simple, ST – simple-to-transitional, T – transitional, TC – transitional-to-complex and C – complex craters. The scaling is denoted by: IS – impactor size scaling, and CP – coupling parameter scaling. Note that the axes scale is linear in (a) and logarithmic in (b).

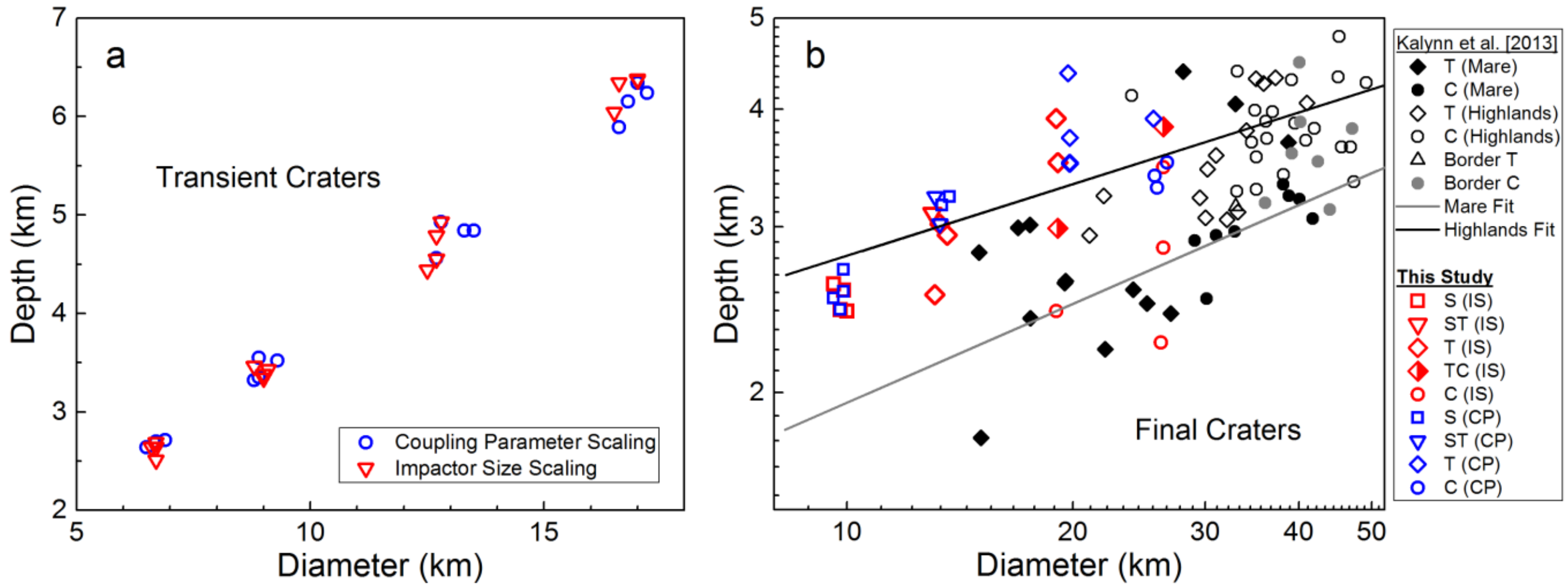

**Figure 3:** The temporal evolution of craters for simulation sets 100 and 1000. The left (a-c) and right (d-e) columns show the comparison between the coupling parameter scaling and the impactor size scaling, respectively. The impact velocities and the simulation numbers are also shown. All resulting craters are simple craters.

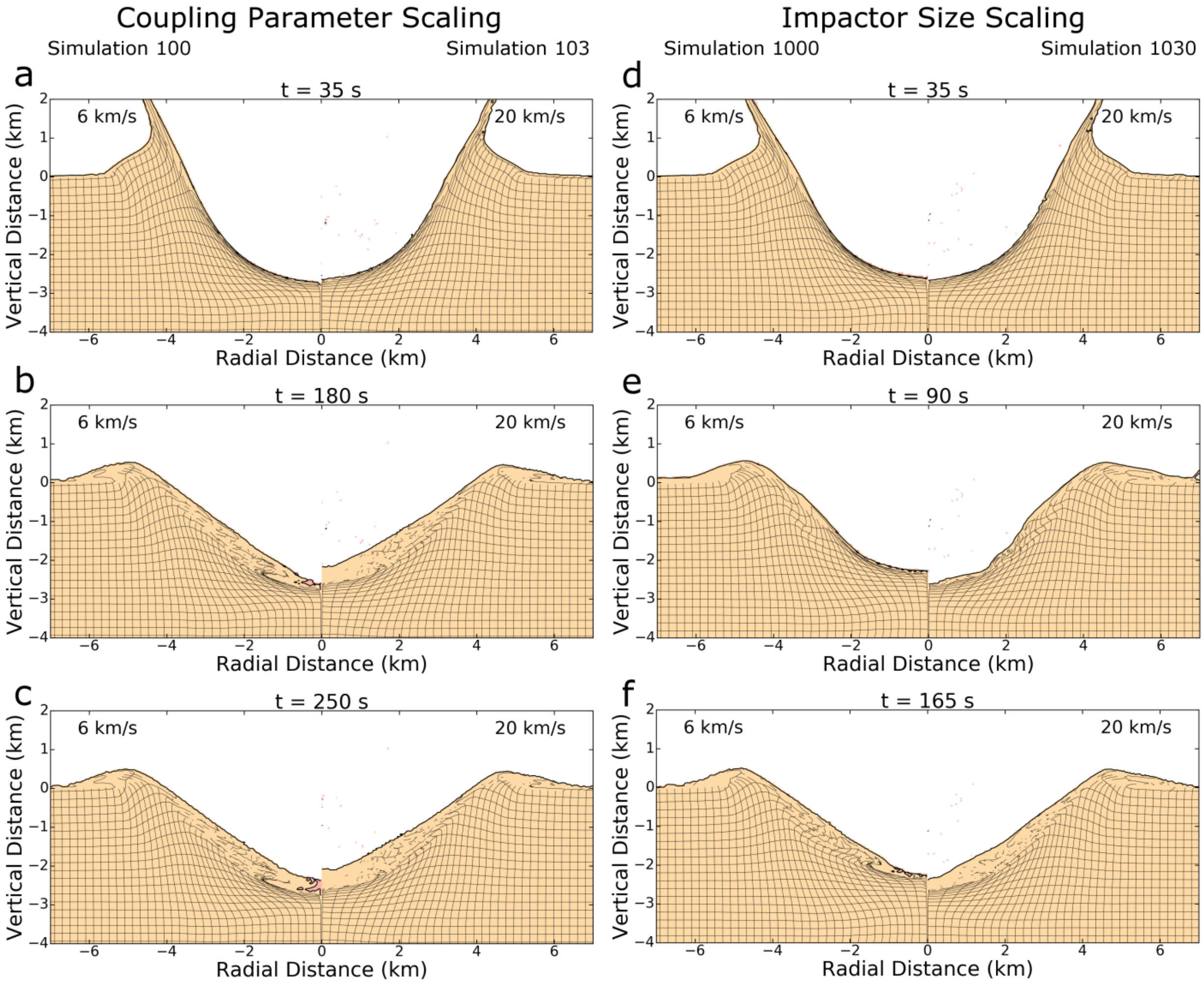

**Figure 4:** The temporal evolution of craters for simulation sets 200 and 2000. The left (a-c) and right (d-e) columns show the comparison between the coupling parameter scaling and the impactor size scaling, respectively. The impact velocities and the simulation numbers are also shown. The resulting craters are simple (simulation 200), simple-to-transitional (simulations 203 and 2030) and transitional (simulation 2000).

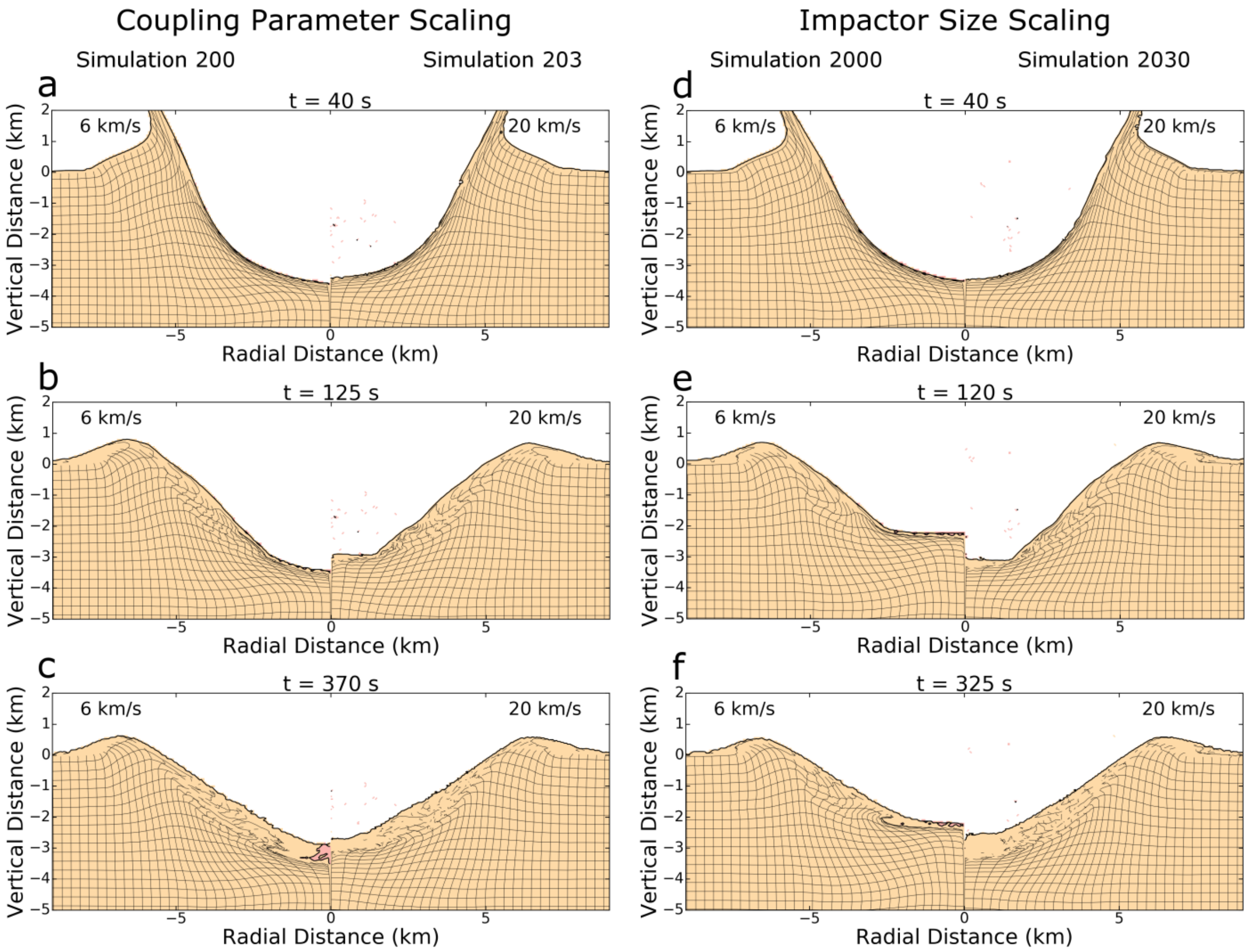

**Figure 5:** The temporal evolution of craters for simulation sets 300 and 3000. The left (a-c) and right (d-e) columns show the comparison between the coupling parameter scaling and the impactor size scaling, respectively. The impact velocities and the simulation numbers are also shown. The resulting craters are mainly transitional, except for simulation 3000, which produces a complex crater.

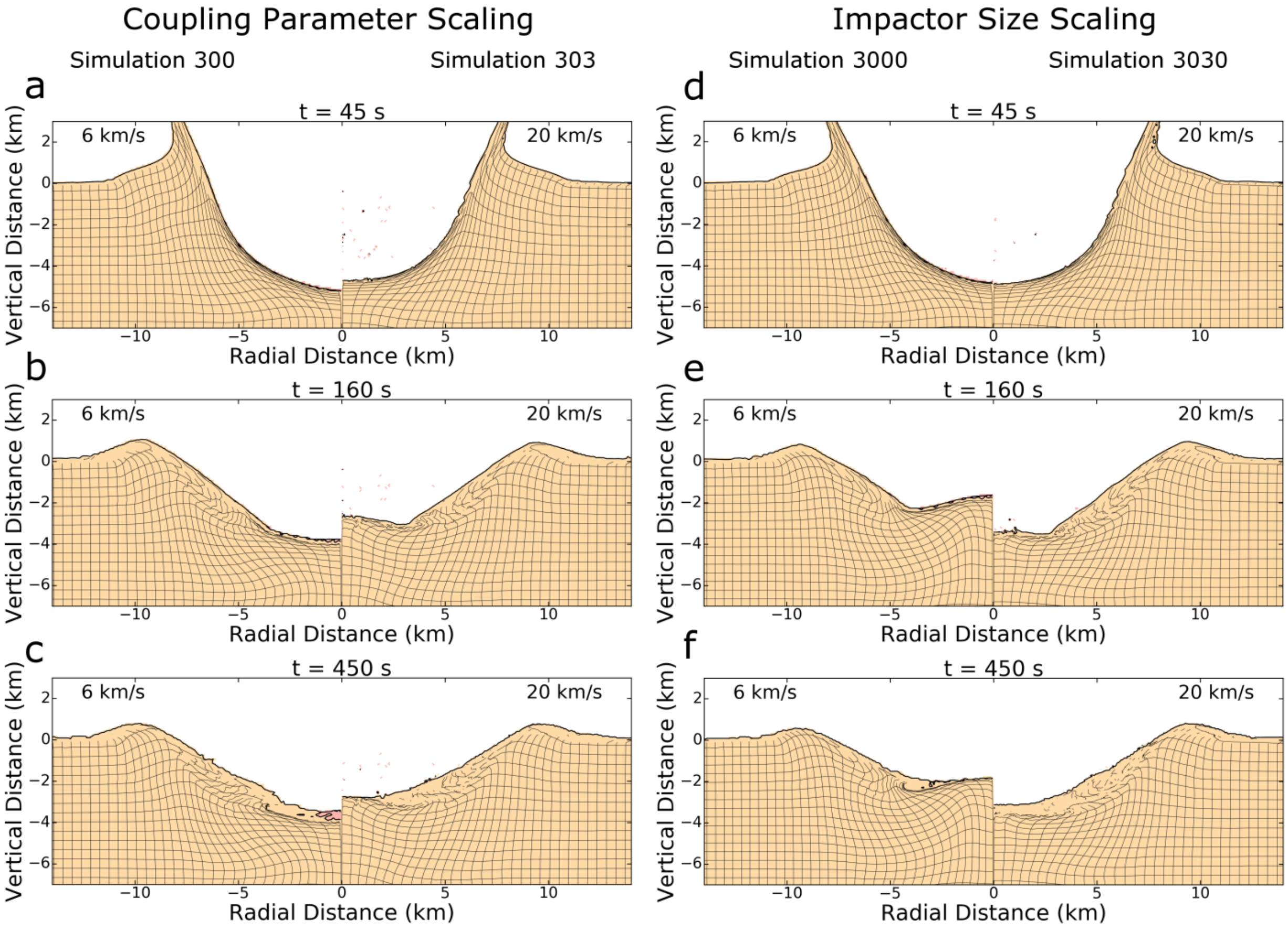

**Figure 6:** The temporal evolution of craters for simulation sets 400 and 4000. The left (a-c) and right (d-e) columns show the comparison between the coupling parameter scaling and the impactor size scaling, respectively. The impact velocities and the simulation numbers are also shown. The resulting craters are transitional (simulation 400), transitional-to-complex (simulation 4030) and complex (simulations 403 and 4000).

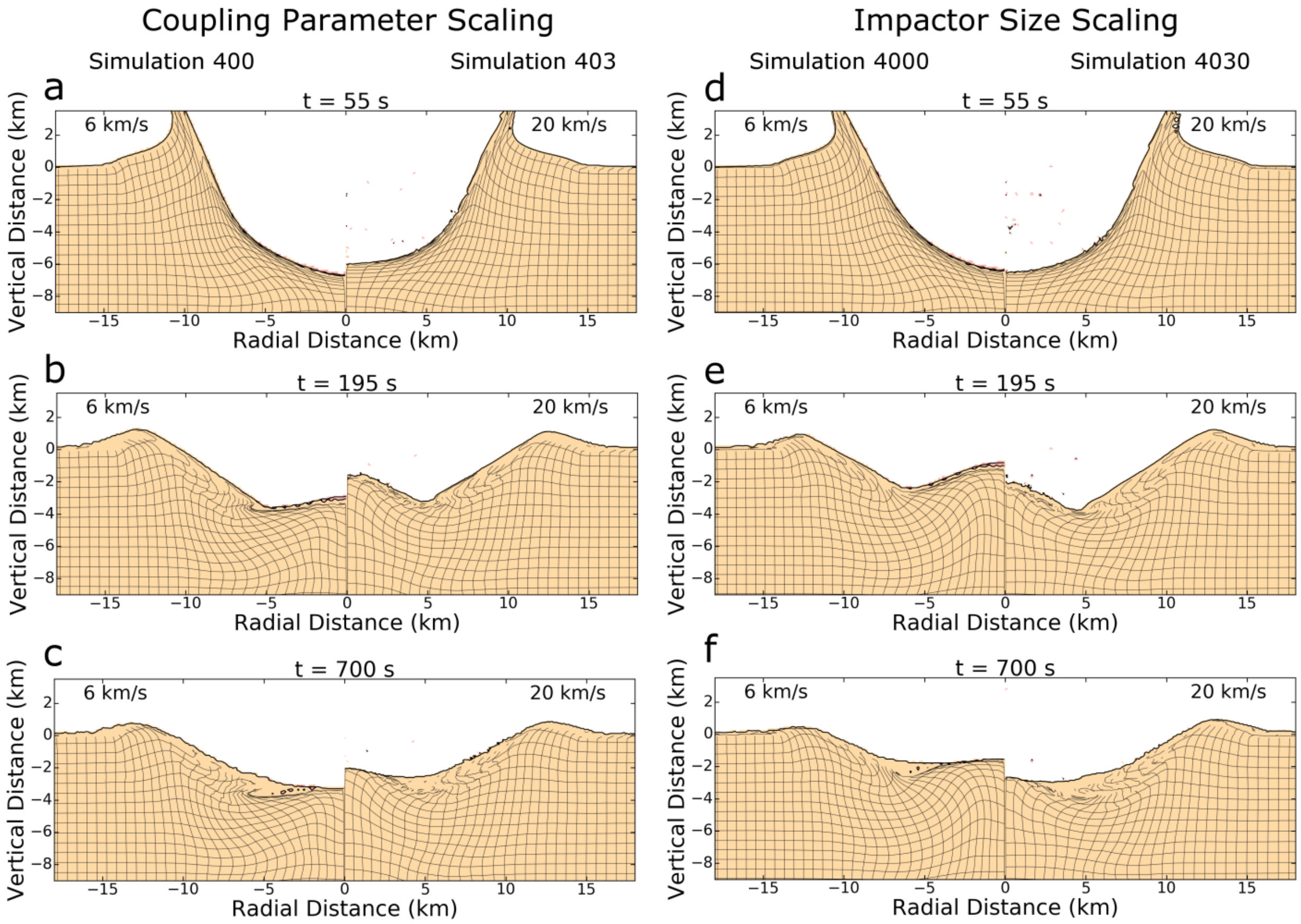

**Figure 7:** The crater profile comparison between real lunar craters and simulated craters. The coupling parameter scaling and the impactor size scaling implementations are shown in panels (a-d) and (e-h), respectively. For clarity, we show the craters produced by large/slow (blue line) and small/fast (red line) impactors only. We note that the morphology and depth profiles at $v_i$ = 15 km/s (not shown in the plot) are identical for both scaling assumptions, since the acoustic fluidization constants are also the same. The impacts at 10 km/s and 15 km/s fall somewhere in-between or very close to those two extremes (see Table 2 for final crater depth measurements).

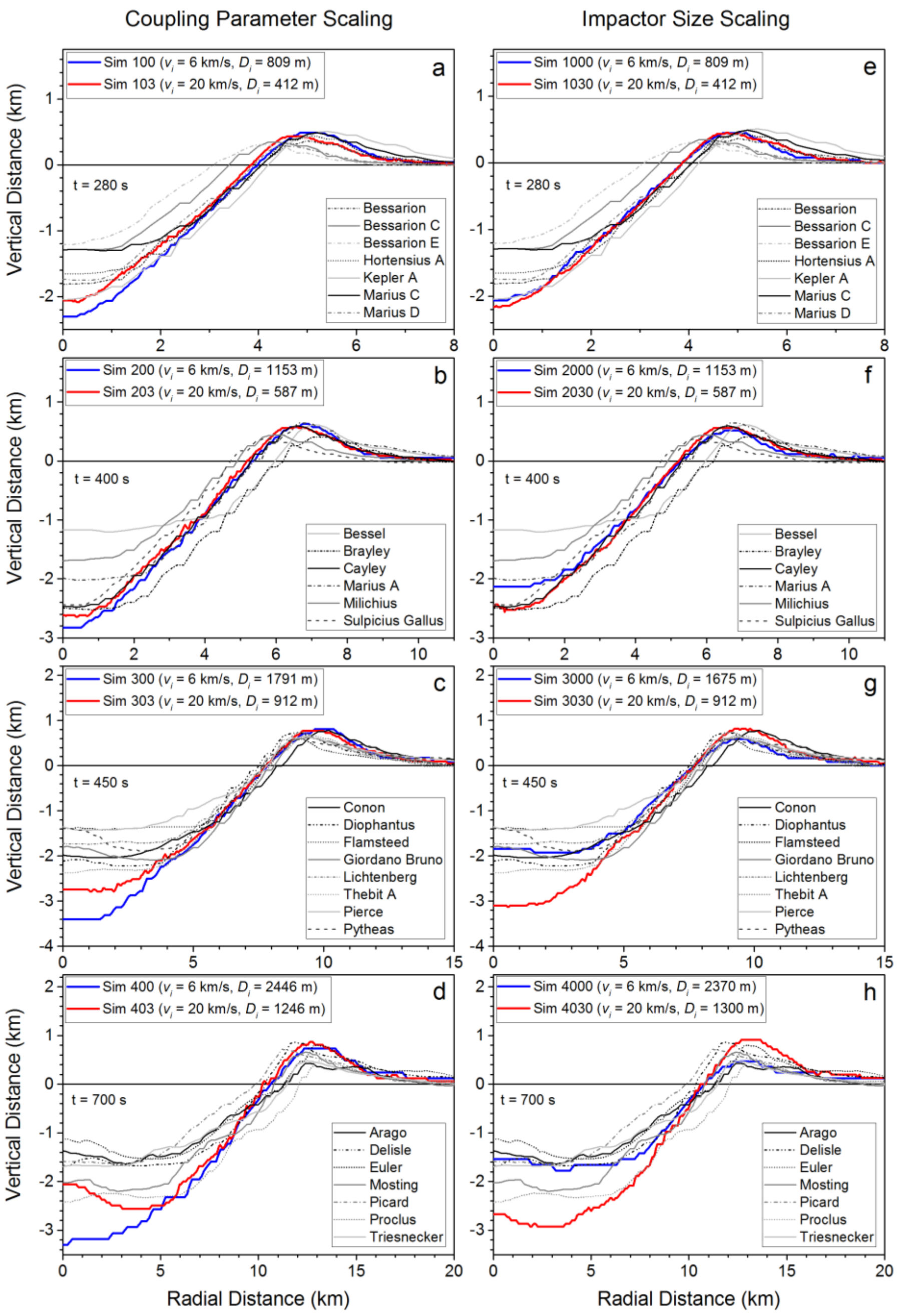

**Figure 8:** The strength of the acoustic fluidization field ($Y_{Ac}$), represented by the ratio of the material strength with the acoustic fluidization included ($Y_d$) to the material strength without the acoustic fluidization ($Y_s$), as: $Y_{Ac} = 1-(Y_d/Y_s)$), for the simulations sets 300/303 (a-c) and 3000/3030 (d-f). At 5 s mark, both large/slow and small/fast impactors produce the acoustically fluidized regions comparable in size (a,d). After 80 s, however, the difference between the two scaling implementations is more evident. When the coupling parameter scaling is used (b), the size of acoustically fluidized zone is relatively the same for large/slow and small/fast impactor (note that the grid resolution differs between these two (see Methodology section)). For the impactor size scaling, the acoustically fluidized zone is larger for the large/slow impactor (e). After 120 s, the effect of acoustic fluidization is nearly dissipated in the coupling parameter scaling (c), while it is still strongly present in the crater produced by the large/slow impactor (f).

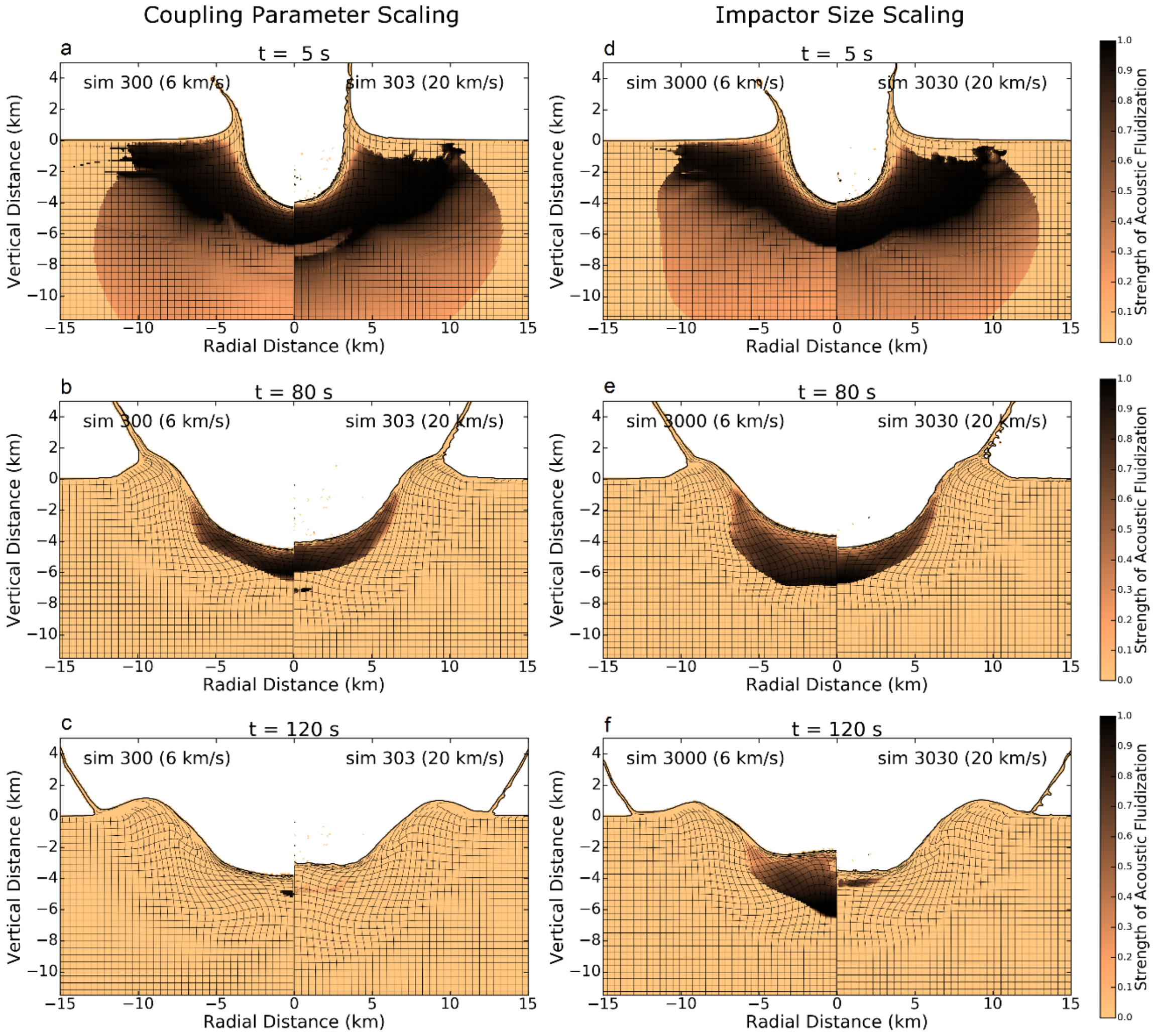